\documentclass[iop]{emulateapj}
\usepackage{natbib, epsfig}

\pdfoutput=1

\newcommand{\um}{\mu{\rm m}}
\shorttitle{UKIDSS Red Quasars}
\shortauthors{Glikman et al.}

\begin{document}


\title{Dust Reddened Quasars in FIRST and UKIDSS: Beyond the Tip of the Iceberg}


\author{Eilat Glikman\altaffilmark{1,2}, Tanya Urrutia\altaffilmark{3}, Mark Lacy\altaffilmark{4}, S.~G. Djorgovski\altaffilmark{5}, Meg Urry\altaffilmark{1}, Scott Croom\altaffilmark{6,7},   Donald P. Schneider\altaffilmark{8,9}, Ashish Mahabal\altaffilmark{5}, Matthew Graham\altaffilmark{5}, Jian Ge\altaffilmark{10}}


\altaffiltext{1}{Department of Physics and Yale Center for Astronomy and Astrophysics, Yale University, P.O. Box 208121, New Haven, CT 06520-8121; email: eilat.glikman@yale.edu}
\altaffiltext{2}{NSF Astronomy and Astrophysics Postdoctoral Fellow}
\altaffiltext{3}{Leibniz Institut f�r Astrophysik, An der Sternwarte 16, 14482 Potsdam, Germany}
\altaffiltext{4}{National Radio Astronomy Observatory, Charlottesville, VA}
\altaffiltext{5}{California Institute of Technology, Pasadena, CA, 91125}
\altaffiltext{6}{Sydney Institute for Astronomy (SIfA), School of Physics, University of Sydney, NSW 2006, Australia}
\altaffiltext{7}{ARC Centre of Excellence for All-sky Astrophysics (CAASTRO)}
\altaffiltext{8}{Department of Astronomy and Astrophysics, The Pennsylvania State University, University Park, PA 16802}
\altaffiltext{9}{Institute for Gravitation and the Cosmos, The Pennsylvania State University, University Park, PA 16802}
\altaffiltext{8}{Astronomy Department, University of Florida, 211 Bryant Space Science Center, PO Box 112055, Gainesville, FL 32611, USA}

\begin{abstract}
We present the results of a pilot survey to find dust-reddened quasars by matching the FIRST radio catalog to the UKIDSS near-infrared survey, and using optical data from SDSS to select objects with very red colors.  The deep $K$-band limit provided by UKIDSS allows for finding more heavily-reddened quasars at higher redshifts as compared with previous work using FIRST and 2MASS.  
We selected 87 candidates with $K\le17.0$ from the UKIDSS Large Area Survey (LAS) First Data Release (DR1) which covers 190 deg$^2$.  These candidates reach up to $\sim 1.5$ magnitudes below the 2MASS limit and obey the color criteria developed to identify dust-reddened quasars.  
We have obtained 61 spectroscopic observations in the optical and/or near-infrared as well as classifications in the literature and have identified 14 reddened quasars with $E(B-V)>0.1$, including three at $z>2$.  
We study the infrared properties of the sample using photometry from the {\it WISE} Observatory and find that infrared colors improve the efficiency of red quasar selection, removing many contaminants in an infrared-to-optical color-selected sample alone.  
The highest-redshift quasars ($z \gtrsim 2$) are only moderately reddened, with $E(B-V)\sim 0.2-0.3$.   We find that the surface density of red quasars rises sharply with faintness, comprising up to $17\%$ of blue quasars at the same apparent $K$-band flux limit.
We estimate that to reach more heavily reddened quasars (i.e., $E(B-V) \gtrsim 0.5$) at $z>2$ and a depth of $K=17$ we would need to survey at least $\sim 2.5$ times more area.
\end{abstract}


\section{Introduction}

The relationship between supermassive black hole (SMBH) growth and the growth of galaxies in the Universe is a major outstanding issue in observational cosmology.  There is ample evidence that SMBHs are somehow linked to their host galaxies. Scaling relations such as the $M_{BH} - \sigma$ relationship \citep{Gebhardt00,Ferrarese00} and $M_{BH} - L_{\rm bulge}$ relationship \citep{Marconi03,Bennert10} suggest that galaxies and their nuclear black holes grow in tandem following a process that is still poorly understood, but which may involve feedback from the SMBH regulating the evolution of its host galaxy.

Merger-driven hierarchical structure formation has been an attractive model for explaining 
the co-evolution of SMBHs and galaxies.  This model was originally motivated by observations of ultra-luminous infrared galaxies (ULIRGs) whose morphologies showed mergers and interaction and which revealed buried, dust-enshrouded quasars as well as high levels of star formation \citep{Sanders88a}.  Numerical simulations of major mergers between galaxies that  host SMBHs also predict a relationship between quasar ignition, which begins in a heavily enshrouded state, and intense star formation induced by the mergers \citep{DiMatteo05,Hopkins06a}.  These simulations reveal the effects of feedback on host galaxy growth, enabling these systems to arrive on the $M-\sigma$ relationship, post merger, and reproduce the bright end of the mass function for galaxies, which is observed to be far steeper than the halo mass function.

Despite the successes of these merger-driven models, recent observations suggest a more complicated picture. \citet{Schawinski11} and \citet{Kocevski12} showed that moderate luminosity ($10^{42} {\rm~erg~ s}^{-1} < L_X < 10^{44} {\rm~erg~ s}^{-1}$)  X-ray-selected AGN at $1.5 < z < 3$ reside in undisturbed, disk-dominated galaxies.
In an analysis of the merger fraction seen in quasar and AGN samples as a function of luminosity, \citet{Treister12} find that mergers dominate only at the highest luminosities (i.e.,  the quasar regime where $L_{\rm bol} \ge 10^{46}$ erg s$^{-1}$).  
 In addition, theoretical models are beginning to include more complicated scenarios for black hole fueling and galaxy evolution; it is possible that stochastic accretion dominates at low luminosities while mergers drive fueling at high luminosities \citep[e.g.,][]{Hopkins06d,Hirschmann12}.

In support of merger-driven co-evolution at the high luminosity end, \citet{Glikman12} identified a population of objects in which merging appears to be the dominant driver of co-evolution and feedback.  
A large population of dust-reddened quasars has been identified by matching the FIRST \citep{Becker95} and 2MASS \citep{Skrutskie06} surveys and selecting objects with very red optical-through-near-infrared colors (we refer to this sample hereafter as F2M; \citealp{Glikman04,Glikman07,Glikman12}; and F2MS; \citealp{Urrutia09}).  Spectroscopic observations of these sources have identified $\sim 120$ red quasars spanning a broad range of redshifts ($0.1 < z \lesssim 3$) and reddenings ($0.1<E(B-V)\lesssim 1.5$).  F2M red quasars are the most luminous objects in the Universe after correcting for reddening, and their fraction increases with increasing luminosity.  They live in merger dominated hosts with elliptical galaxy profiles \citep{Urrutia08}.  Their spectra show high fraction of LoBAL and FeLoBAL features providing evidence for outflows that could be associated with feedback \citep{Urrutia09,Farrah12,Glikman12}.  Many have extremely high accretion rates, and the high Eddington-ratio systems have large bulge luminosities relative to their black hole masses, suggesting that their stars have formed before the black-hole has finished growing \citep{Urrutia12}.  This body of evidence suggests that the dust-reddened quasars in the F2M survey are systems in which a merger-fueled, heavily obscured quasar is emerging from its shrouded environment.  Based on the statistical frequency of these sources, compared with optically-selected, blue quasars, \citet{Glikman12} estimate that the duration of the red phase is $\sim 20\%$ as long as the unobscured quasar lifetime.  

In addition to the F2M sample, there have been several efforts to identify populations of red quasars in the radio \citep[e.g.,][]{Webster95,White03b} and mid-infrared \citep[e.g.,][]{Lacy04,Lacy07,Polletta06,Polletta08,Stern12}.   \citet{Warren00} developed a technique exploiting the $K$-band excess in the power-law shape of quasar spectra compared with stars ('KX' selection method) that is less biased than the optical to dust-extinction.  \citet{Maddox08} and \citet{Maddox12} have utilized the KX method to identify quasar samples including moderately reddened sources out to $E(B-V) < 0.5$.
Recently, a targeted search for heavily reddened quasars at $z\sim 2$ using near-infrared selection with no requirement for radio detection criterion discovered 12 red quasars that show similar properties to the F2M quasars \citep{Banerji12,Banerji13}.  They arrived at the same interpretation as \citet{Glikman12} and \citet{Urrutia12}:  dust reddened quasars are a transitional phase in quasar-galaxy co-evolution.  

The F2M survey was relatively shallow, and only revealed the ``tip of the iceberg'' for reddened AGN; at higher redshifts ($z > 1.5$) only the most intrinsically luminous objects are seen.  To reach the heavily-reddened higher-redshift analogs to the F2M quasars, a more sensitive near-IR survey is needed to tease out the luminosity and redshift dependences of red quasars.

The ideal survey for extending the F2M red quasar survey is the UKIRT Infrared Deep Sky Survey \citep[UKIDSS;][]{Lawrence07}. UKIDSS is a near-IR imaging sky survey comprised of five tiered surveys with varying depths and areas to supplement the wavelength coverage of the sky beyond the optical. The largest of these, the Large Area Survey (LAS) has, to date, covered $\sim3000$ deg$^2$ in the $Y$, $J$, $H$, and $K$ bands down to $K \sim18$ magnitudes, which is approximately 2.5 magnitudes deeper than 2MASS.  In addition, the image quality of UKIDSS is comparable to optical CCD-based surveys, with a typical full-width at half-maximum (FWHM) of the point spread function (PSF) below an arc-second (compared with FWHM of 2\arcsec\ for 2MASS point sources).

In this paper, we present a sample of red quasars using the UKIDSS survey down to $K=17$, or $\sim 1.5$ magnitudes fainter than 2MASS in the near-infrared.  We construct our sample by matching the FIRST radio survey to the UKIDSS First Data Release \citep[DR1][]{Warren07} and use optical photometry from the Sloan Digital Sky Survey \citep[SDSS;][]{York00} to select objects with red optical-to-near-infrared colors.  

The paper is organized in the following manner.  In Section 2 we describe out color selection technique and construction of the red quasar candidate list.  We describe our spectroscopic follow-up observations in Section 3.  In Section 4, we discuss other methods of selecting reddened quasars, their advantages and drawbacks.  We analyze the surface density and demographics of this deeper red quasar sample in Section 5.  In Section \ref{sec:ebv} we use all available photometric and spectroscopic data for our quasars to estimate the reddening experienced by each source and conclude our findings in Section 7.  When calculating distances, luminosities and other cosmology-dependent quantities, we use the parameters: $H_0$ = 70 km s$^{-1}$ Mpc$^{-1}$, $\Omega_M = 0.30$, and $\Omega_\Lambda = 0.70$.

\section{Red Quasar Color Selection}

\defcitealias{Urrutia09}{F2MS}

In this paper we explore the space density of red quasars extending $\sim 1.5$ magnitudes below  the 2MASS $K$-band flux limit by applying selection criteria similar to the F2M survey.  
\citet{Glikman04} found that the colors $R-K>4$ and $J-K>1.7$ were efficient color cuts for finding red quasars, which were used to identify 120 red quasars in \citet{Glikman07} and \citet{Glikman12}.  However, the F2M survey used optical magnitudes from the Guide Star Catalog 2.2 \citep{Lasker08}, which are based on digitized scans of the POSS-II photographic plates \citep{Reid91}.  The $R$-magnitude used in the $R-K$ color is therefore not equivalent to the SDSS $r$-band.
\citet{Urrutia09} used a combination of FIRST, 2MASS and SDSS to create a sample of red quasars (hereafter, we refer to this sample as F2MS).  The areal coverage of the F2M and F2MS surveys overlapped and they have many sources in common.  However, \citet{Urrutia09}, required $r-K>5$\footnote{$r-K>5$ is roughly equivalent to $R-K>4.5$ \citep{Windhorst91}. } and $J-K>1.3$\footnote{The SDSS magnitudes are on the AB magnitude system \citep{Oke83}, while 2MASS and UKIDSS report Vega magnitudes.  These color cuts represent colors computed directly from the respective databases, with no corrections/transformations applied.}.  We found, when comparing the full set of F2M quasars to the subset selected by \citet{Urrutia09} that five quasars with   $1.5<J-K<1.7$ were missed by the F2M color cuts, which amounted to 9\% of the \citet{Urrutia09} sample.  In order to remedy this incompleteness, we amend the original F2M color cuts to take into account the SDSS filters and the incompleteness in $J-K$; for this pilot study of heavily reddened quasars in FIRST+UKIDSS DR1, we require $r-K > 5$, and $J-K > 1.5$ with $K<17$.  
 
\begin{figure}
\epsscale{1.2}
\plotone{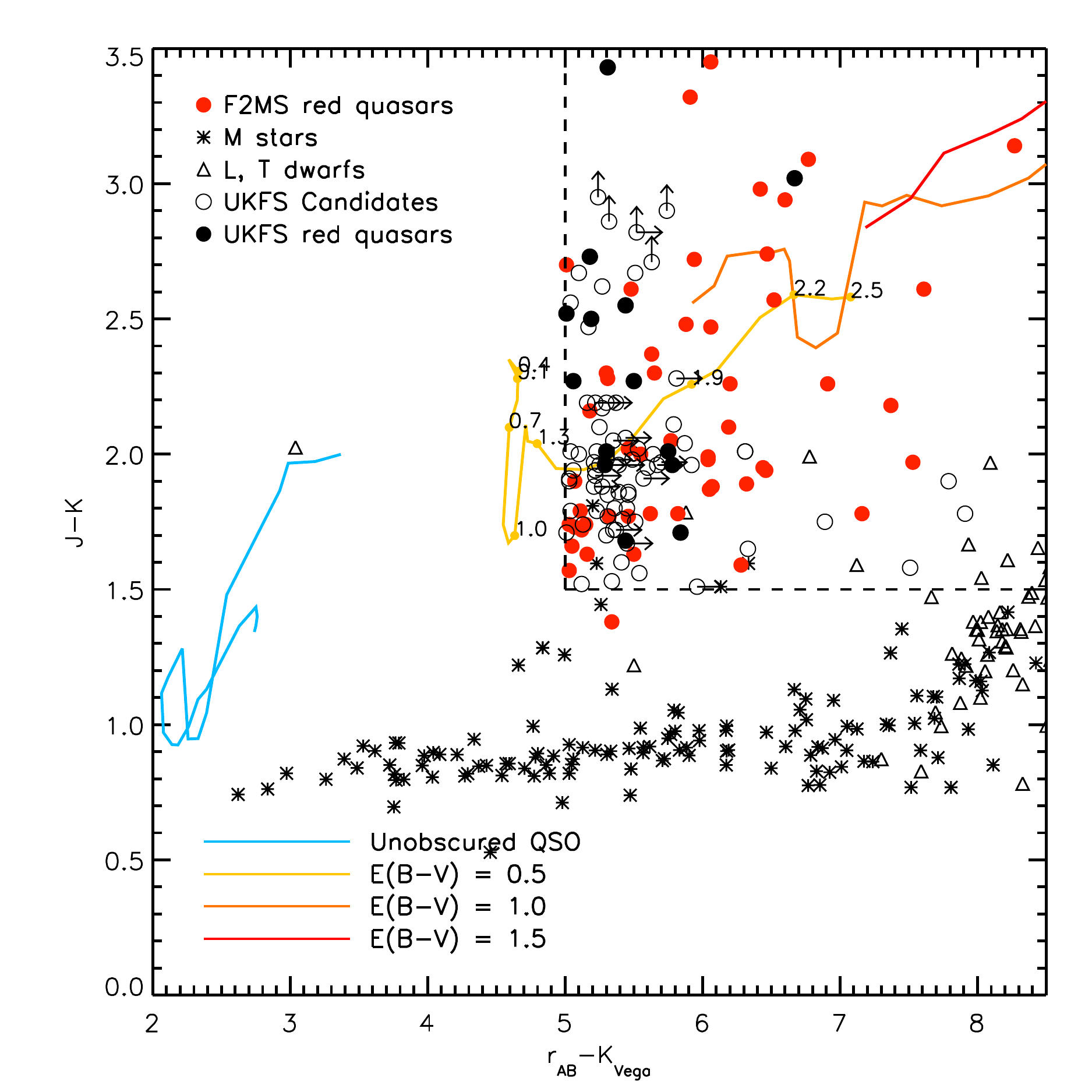}
\caption{Colors of quasars with various amounts of reddening and cool field stars, in  $J-K$ vs.~$r-K$.  The solid lines are modeled Type I quasar colors between $z=0.1$ to $2.8$. The blue line shows the colors of an average quasar with no reddening, while the yellow, orange and red lines show the color tracks of quasars reddened by $E(B-V) = 0.5, 1.0$ and 1.5 magnitudes with an SMC extinction law, respectively.  Redshifts spaced by $\Delta z = 0.3$ are labeled on the yellow ($E(B-V)=0.5$) line. 
For consistency with the format in which the data products are made available, we use AB magnitudes for the SDSS-$r$ band and Vega magnitudes for the near-infrared $J$ and $K$ bands.
Overplotted are the red quasars from \citet{Urrutia09} (red circles), all but one of which fall into our $J-K=1.5$ and $r-K=5$ color cuts.  
Our color cuts (dashed lines) mostly avoid low mass stars (black asterisk, triangle symbols), which are further excluded by our radio selection.
We overplot the UKFS candidates with open circles and the confirmed red quasars with filed circles.
Because of the $r-K$ color cut, our color selection may miss low-redshift, mildly-reddened sources.}\label{fig:color}
\end{figure}

Figure \ref{fig:color} shows a series of modeled quasar colors out to $z\le 2.5$ with various amounts of reddening.  The yellow line with $E(B-V) = 0.5$ is labeled with the modeled quasars' redshifts.  The confirmed F2MS red quasars from \citet{Urrutia09} are shown with red circles.  We also plot the colors of M, L and T dwarfs (asterisks and triangles, respectively) which our $J-K>1.5$ color cut largely avoids\footnote{The surface density of low mass stars overwhelms that of red quasars in infrared surveys, thereby compelling us to add the radio selection to help eliminate stellar contamination from our sample.}.  \citet{Glikman12} demonstrated that added light from a host galaxy does not signficantly affect the color of these reddened quasars; the largest consequence of the $r-K > 5$ color cut is the possibility of missing lightly reddened, lower-redshift ($z \lesssim 1.3$, $E(B-V) \lesssim 0.5$) quasars.

We matched the 3 April, 2011 FIRST radio catalog\footnote{\tt http://sundog.stsci.edu/first/catalogs/readme\_03apr11.html} \citep{White97} to the UKIDSS First Data Release \citep[DR1;][]{Warren07}.  This yielded a matched catalog of 4890 sources (2432 with $K\le17$), including all UKIDSS matches within 2\arcsec\ of a FIRST source, not just the nearest match.  We make no restriction on the UKIDSS morphological classification in our selection, including all {\tt mergedClass} values from ``stellar'' to ``galaxy''.  Most of the objects ($87\%$) are classified as galaxies ({\tt mergedClass = 1}) with only a small fraction ($8\%$) classified as stellar ({\tt mergedClass = -1}).  The remaining sources are classified as either {\tt probableStar}, {\tt probablyGalaxy}, {\tt noise} or {\tt saturated} \citep[see Appendix A of ][]{Dye06}.  

In order to obtain their optical magnitudes, we then matched these objects, using the FIRST position, to the SDSS DR6 \citep{Adelman-McCarthy08} catalog with a search radius of 2\arcsec.   There were 4000 FIRST+UKIDSS sources with a match to SDSS, 2465 of which have $K\le17$ (including multiple matches to the same radio source).   \citet{Glikman12} showed that $\sim 50\%$ of the F2M red quasars were classified as extended ({\tt type = 3}) in SDSS.  To avoid any morphological bias we make no eliminations based on morphology and keep all classifications from the SDSS.   At this stage, the majority of FIRST+UKIDSS+SDSS sources are classified as extended in the optical ($84\%$ versus $16\%$ with a stellar classification).

We also include FIRST+UKIDSS sources with $K\le17$ that are not detected in SDSS, as this sample is likely to contain the most heavily reddened quasars.  There were 901 FIRST+UKIDSS sources with no match in the SDSS catalog within 2\arcsec. The SDSS DR6 does not include the deeper observations over Stripe 82, a region of the SDSS footprint covering the area $-50\degr <  \alpha_{2000}<59\degr$, $-1.25\degr < \delta_{2000} < 1.25\degr$ that has been re-visited approximately 80 times and whose co-added frames reach $\sim 2$ magnitudes deeper than the nominal SDSS magnitude limits.  Although there is considerable overlap between the UKIDSS DR1 (fields LAS 5, LAS 6, LAS 7 and LAS 8) and Stripe 82, we did not incorporate the magnitudes from the co-added Stripe 82 data in our source selection, to maintain uniformity in our selection process.  

We applied the color cuts shown in Figure \ref{fig:color} to select red quasar candidates.  For the optically undetected sources, we used the quoted 95\% completeness magnitude limits for SDSS, $r = 22.2$ and $i=21.3$ \citep{Adelman-McCarthy06} when computing colors.  This color limit automatically includes all undetected sources as candidates, since all the candidates have $K\leq17$, which implies that  all optically undetected sources have $r-K \ge 5.2$.   

We extracted UKIDSS image cutouts of all the sources using the Wide Field Camera \citep[WFCAM;][]{Casali07} Science Archive \citep[WSA;][]{Hambly08} as well as cutouts from SDSS and removed objects that appeared to be image artifacts, e.g., artifacts associated with nearby bright stars, cross talk and, false detections due to imperfect sky subtraction near the edges of a field.  The final quasar candidate list of FIRST+UKIDSS sources with $K\le17$ (with and without SDSS matches) obeying the color criteria $r-K>5$ and $J-K>1.5$ contains 87 objects. We call this the UKFS candidate catalog, listed in Table 1, which includes 69 sources with SDSS detections and 18 candidates without SDSS detections.

Since the versions of FIRST and SDSS that we use fully overlap the UKIDSS DR1 area, the size of our survey is determined by the UKIDSS footprint which is 189.6 deg$^2$ \citep{Warren07}.  We use this area to determine the surface density and demographics of the fainter red quasars found in this survey.  

\section{Observations}

We obtained 61 spectroscopic identifications of our 87 candidates.  These identifications were primarily determined from forty-four near-infrared spectra acquired with the TripleSpec spectrograph \citep{Herter08} on the Palomar Hale telescope during five observing runs between August 2008 and April 2013.  The TripleSpec data were reduced using a modified version of the Spextool software, which includes flat-fielding, sky-subtraction, co-addition of individual frames, extraction and wavelength calibration using sky lines \citep{Cushing04}.  We also obtained a spectrum of a nearby A0V star at a similar airmass (aiming for $\Delta$ airmass $<0.1$ between the target and the star) after each object.  The A0V spectrum is used to correct for telluric absorption \citep{Vacca03}

An additional fifteen optical spectra were obtained from the SDSS.  Of these, eleven are from the most recent Data Release 9 \citep[DR9;][]{Ahn12} public spectroscopic database, which includes spectroscopy from the Baryon Oscillation Spectroscopic Survey \citep[BOSS;][]{Eisenstein11}.  BOSS spectra are taken with a fiber-fed, mulit-object spectrograph \citep{Smee12}.  A pipeline reduces, classifies and assigns redshifts to the spectra \citep{Bolton12}.

Eleven spectra came from the AAOmega-UKIDSS-SDSS Survey (AUS; Croom et al. in prep) in the Stripe-82 region, an additional three spectra were obtained with the Low Resolution Imaging Spectrograph \citep[LRIS;][]{Oke95} on the W. M. Keck telescope on 15 October 2012, and one object was identified as a luminous red galaxy (LRG) by the 2dF-SDSS LRG and QSO survey \citep[2SLAQ;][]{Cannon06}.  In addition, two sources have photometric redshifts determined from the Red-Sequence Cluster Survey \citep{Hsieh05}.  And we re-discover two F2M red quasars from \citet{Glikman07} and \citet{Glikman12}.  Column (11) of Table 1 lists the origin of the spectroscopy for each candidate in our sample.  

\begin{figure}
\epsscale{1.2}
\plotone{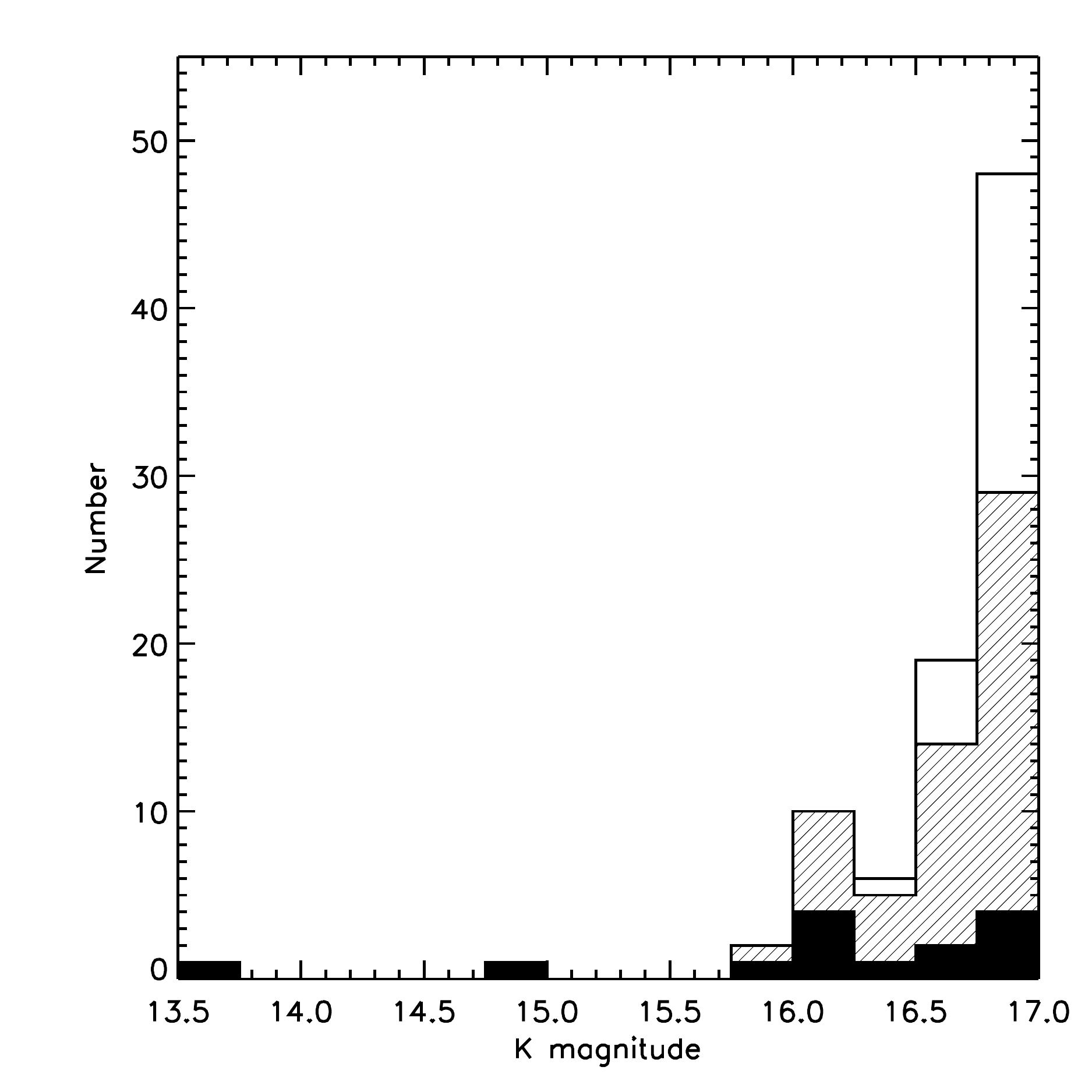}
\caption{$K$-magnitude distribution for the UKFS candidates binned by 0.25 magnitudes.  The open bars show all 87 candidates.  The shaded histogram shows all 63 sources with spectroscopic observations, while the filled histogram shows the 14 confirmed quasars.  Note that the survey is 95\% spectroscopically complete for $K<16.5$.}\label{fig:hist}
\end{figure}

Figure \ref{fig:hist} shows the distribution of $K$-band magnitudes for our candidates.  The shaded histogram shows shows the objects with spectra; confirmed quasars are shown in the filled histogram.   We are 95\% spectroscopically complete below $K=16.5$.  Our completeness drops to 64\% for $16.5 < K < 17$. We recover the two F2M quasars that overlap this area and find an additional 12 quasars, for a total of 14 red quasars in the FIRST+UKIDSS DR1 overlap.  We hereafter refer to these objects as the UKFS (UKIDSS+FIRST+SDSS) red quasar sample.  

In Figure \ref{fig:spec1} we present a spectral atlas of the 14 UKFS quasars in decreasing redshift order.  We label the positions of typical prominent quasar emission lines (Ly$\alpha$~1216, \ion{N}{5}~1240, \ion{Si}{4}~1400, \ion{C}{4}~1550, \ion{C}{3}]~1909, \ion{Mg}{2}~2800, [\ion{O}{2}]~3727, H$\delta$~4102, H$\gamma$~4341, H$\beta$~4862, [\ion{O}{3}]~4959, 5007, H$\alpha$~6563, \ion{He}{1}~10830, Pa$\gamma$~10941, Pa$\beta$~12822~\AA) with vertical dotted lines.  We also plot with a red line the best-fit reddened quasar template \citep[from][]{Glikman06} to the spectra (see Section \ref{sec:ebv} for further discussion).

\begin{figure*}
\plotone{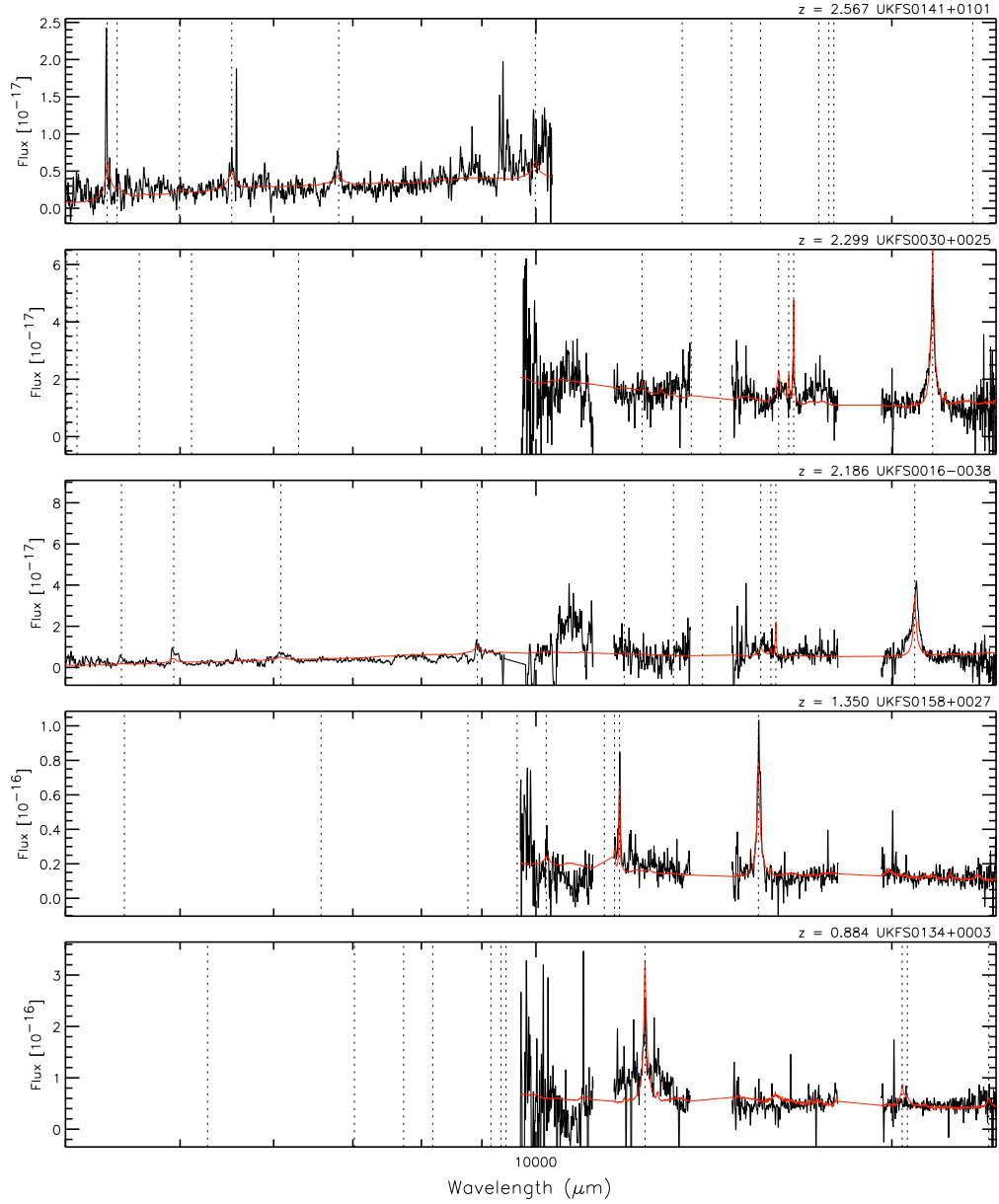}
\caption{Optical and/or near-infrared spectra of UKFS quasars ordered by redshift.  The red line shows the best-fit reddened quasar template to the combined optical and near-infrared spectra.  Typical quasar emission lines are marked with vertical dashed lines: 
Ly$\alpha$~1216,
N~V~1240,
Si~IV~1400,
C~IV~1550,
C~III]~1909,
Mg~II~2800,
[O~II]~3727,
H$\delta$~4102,
H$\gamma$~4341,
H$\beta$~4862,
[O~III]~4959,
[O~III]~5007,
H$\alpha$~6563,
He~I~10830,
Pa$\gamma$~10941,
Pa$\beta$~12822\AA.}\label{fig:spec1}
\end{figure*}

\begin{figure*}
\figurenum{3b}
\plotone{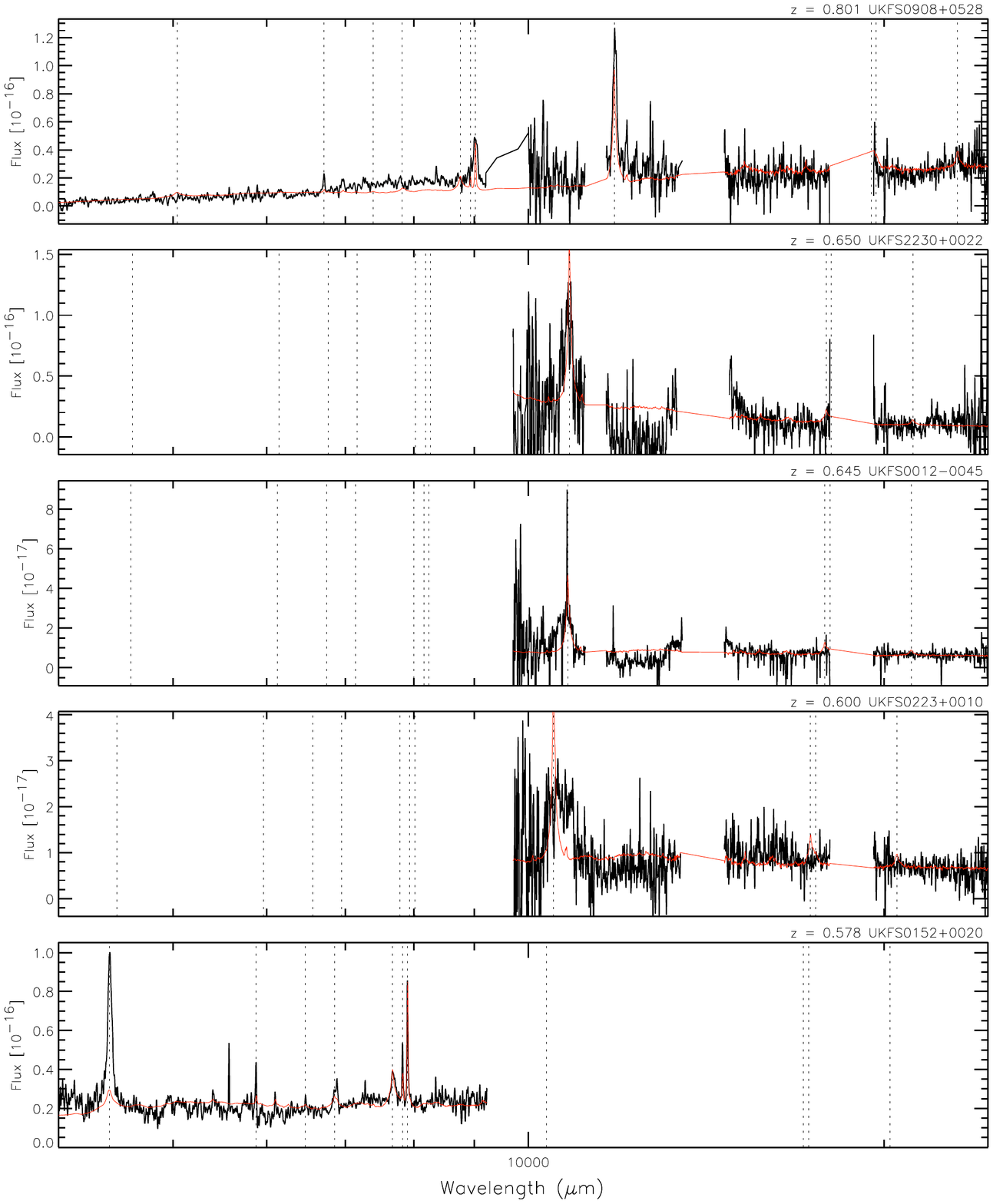}
\caption{{\em Continued.} Optical-through-near-infrared spectra of UKFS quasars. }\label{fig:spec2}
\end{figure*}

\begin{figure*}
\figurenum{3c}
\plotone{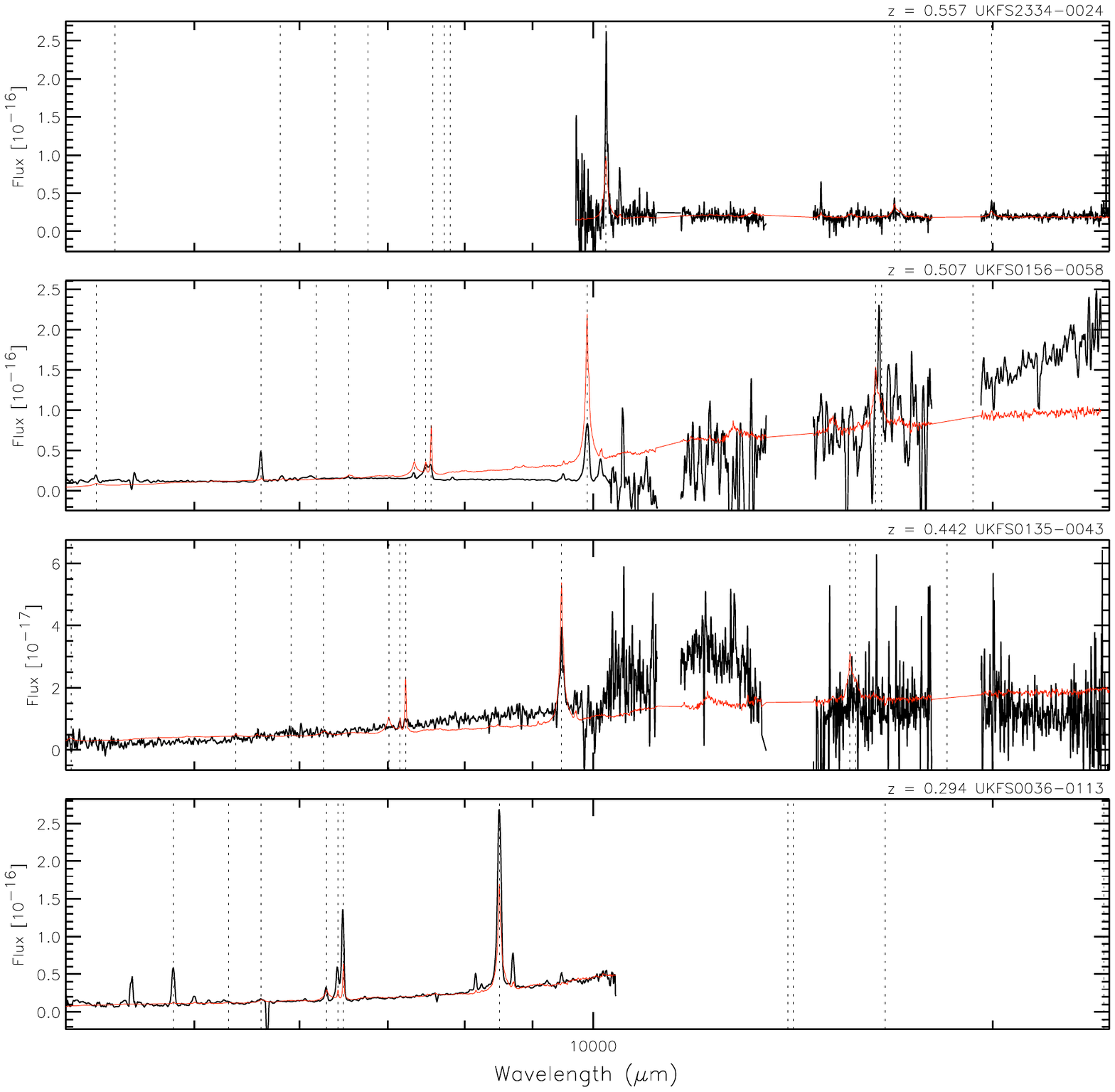}
\caption{{\em Continued.} Optical-through-near-infrared spectra of UKFS quasars. }\label{fig:spec3}
\end{figure*}

\section{Complementary Red Quasar Selection Methods}
\subsection{The KX-Selection}

Because of the power-law nature of a quasar's SED, quasars are separable from stars, whose spectra are blackbodies, in color-color space.  At short wavelengths, quasars are bluer than the bluest stars, giving rise to the so-called ultraviolet excess (UVX).  This feature has been exploited for quasar selection in optical surveys, \citep[e.g.,][]{Sandage65,Schmidt83} resulting in a literature of $\gtrsim 10^5$ spectroscopically confirmed quasars out to $z\sim 2.5$ \citep[e.g.,][]{Veron-Cetty10}.  At long wavelengths the colors of quasars also diverge from stars, appearing redder, giving rise to the so-called $K$-band excess \citep[KX;][]{Warren00}, which allows for efficient quasar selection in the near-infrared.  In addition, since dust extinction is an inverse exponential function of wavelength, near-infrared emission is less affected than the optical and rest-frame UV.  This means that KX selection of quasars is far less sensitive to dust, yet remaining as, or possibly more, efficient than UVX selection.  

In particular, UVX selection fails for $z\gtrsim2.2$ when the Lyman-$\alpha$ line shifts into the $B$-band reddening the observed quasar's $U-B$ color.  At higher redshifts ($z \gtrsim 4$), absorption from the Lyman-$\alpha$ forest further reddens a quasar's UV and optical colors, making UVX selection ineffective and requiring other color selection methods for identifying high redshift quasars \citep[e.g.,][]{Kennefick95,Fan99b,Glikman10}. The KX-selection method extends the redshift range for finding quasars as a result of two effects: (1) optical/UV wavelengths are more susceptible to dust extinction, which is exacerbated at higher redshifts as the rest frame wavelengths are shifted blueward, and (2) Lyman-$\alpha$ forest absorption is not an issue until $z\sim 3.5$ allowing for access to quasars in the $z\sim 2-3$ regime, where optical selection is most incomplete \citep{Warren00,Richards02,Maddox12}.

There have been recent efforts to recover missed quasars, including reddened ones, using KX selection with UKIDSS and SDSS.  These studies have identified known quasars from, e.g., SDSS and other optical surveys, as well as additional quasars with unusual properties and/or at redshifts inaccessible to optical selection.  \citet{Maddox12} used KX-selection to find quasars in the UKIDSS DR4 LAS data combined with SDSS DR7.  This selection resulted in recovering 3294 SDSS quasars, plus 324 new quasars.  To compare the UKFS quasars with the KX-selected sources, we matched Tables 4 and 6 from \citet{Maddox12} to the 16 February 2012 release of the FIRST radio catalog, so that both samples are radio-detected down to the same flux density limit.  There are 263 FIRST matches to the KX-selected, SDSS-identified quasars \citep[Table 6 in][]{Maddox12} and only 9 FIRST matches to the newly discovered KX-selected quasars listed in Table 4 of \citet{Maddox12}.  There is incomplete areal overlap between the two surveys, accounting for some of the missed quasars.  

\begin{figure}
\epsscale{1.2}
\plotone{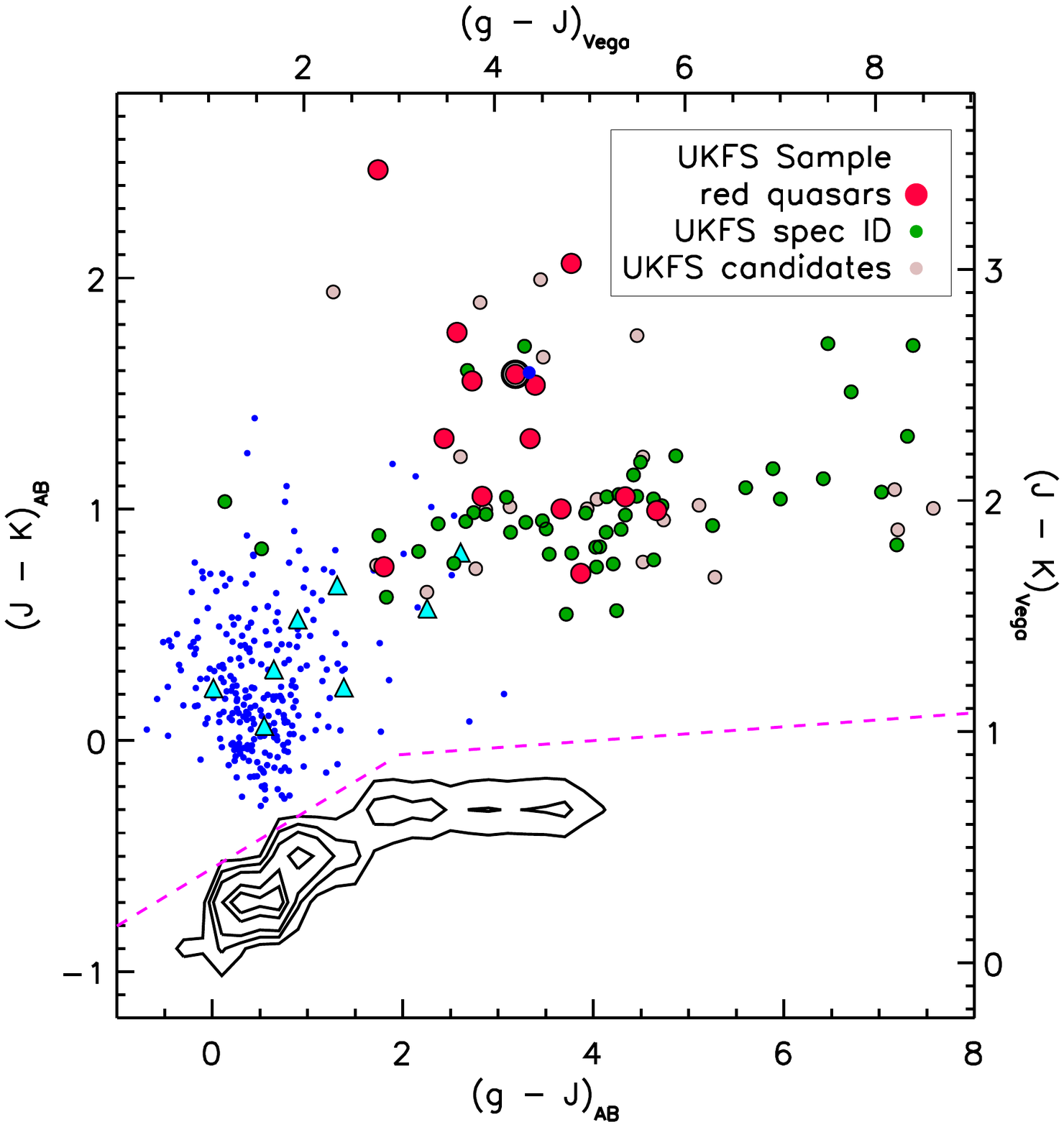}
\caption{The location of the UKFS candidates in KX color space ($J-K$ vs.\ $g-J$) is shown.  We plot quasars found by \citet{Maddox12} with matches in FIRST with blue circles (their recovered quasars from optically selected methods) and cyan (their new quasars).  The colors of sepctroscopically-confirmed stars from SDSS are plotted with contours showing that the KX-selection selection boundary (magenta dashed line) effectively separates quasars from stars.  Our UKFS sample targets the reddest sources. Red circles represent  confirmed quasars, green circles are spectroscopically observed objects showing no broad lines inter spectra, and gray circles are UKFS candidates without spectroscopic observations.  }\label{fig:kx}
\end{figure}

Figure \ref{fig:kx} shows the location of the FIRST-detected quasars identified by \citet{Maddox12} in $g-J$ vs. $J-K$ color-color space.  Since the SDSS photometry is on the AB magnitude system \citep{Oke83}, while UKIDSS uses the Vega standard, \citet{Maddox12} shift the SDSS magnitudes to the Vega system for consistency. We have chosen to use the AB magnitudes since they are naturally representative of physical units (i.e., flux density) without prior knowledge of photometric zero points.  We plot the quasars from  \citet{Maddox12} with blue circles and cyan triangles.  We show the location of spectroscopically confirmed stars from SDSS matched to UKIDSS with black contours, while the magenta dashed line indicates the KX-selection boundary defined in \citet{Maddox08} which separates quasars from stars:
\begin{eqnarray}
%% Vega
(g-J)_{\rm Vega} = 4(J-K)_{\rm Vega} - 0.6
\end{eqnarray}
for  $(J-K)_{\rm Vega} \le 0.9$ and 
\begin{eqnarray}
(g-J)_{\rm Vega} = 33.33(J-K)_{\rm Vega} -27 
\end{eqnarray}
for $(J-K)_{\rm Vega} > 0.9$. The conversions from Vega to AB for the UKIDSS bands are $J_{\rm V} = J_{\rm AB} - 0.938$ and $K_{\rm  V} = K_{\rm AB} - 1.9$; thus the equation for the KX boundary in the AB magnitude transforms to:
\begin{eqnarray}
%% AB
(g-J)_{\rm AB} = 4(J-K)_{\rm AB} - 2.21 
\end{eqnarray}
for $(J-K)_{\rm AB} \le 0.062$ and
\begin{eqnarray}
(g-J)_{\rm AB} = 33.33(J-K)_{\rm AB} + 4.03
\end{eqnarray}
for $(J-K)_{\rm AB} > 0.062$.  

The top and right-hand axes of Figure \ref{fig:kx} are shifted to show colors on the Vega system, for ease of comparison with Figure 2 of \citet{Maddox12}.  In addition to these color cuts, \citet{Maddox12} require that their candidates appear stellar in the UKIDSS images -- a criterion not imposed by the UKFS sample selection.

The UKFS sources are also plotted in Figure \ref{fig:kx}.  Spectroscopically confirmed red quasars are red circles,   spectroscopically observed objects that show no broad emission lines are plotted with green circles, and UKFS candidates with no spectroscopic observations are colored gray.  

Despite 100\% areal overlap between the UKFS and KX surveys, as well as overlapping color criteria, only one UKFS quasar was found by \citet{Maddox12}: UKFS0016$-$0038, which is one of their new quasars \citep[this object is also found in the sample of][]{Banerji12}.  This object is indicated in Figure \ref{fig:kx} with a red circle emphasized by a thick black border, at $(g-J)_{\rm AB} = 3.18$ and $(J-K)_{\rm AB} = 1.58$ representing its colors in the UKFS survey, which derives its magnitudes from the UKIDSS DR1 LAS merged catalog.  The same object appears as a small blue point with slightly different colors, at $(g-J)_{\rm AB} = 3.33$ and $(J-K)_{\rm AB} = 1.59$, which come from the UKIDSS DR4 LAS merged catalog.  

Our UKFS survey finds red quasars missed by the KX survey of \citet{Maddox12}, partly because that survey had a flux limit of $K\le16.6$ and partly because the selection focused on finding sources with $1.0 \le z \le 3.5$ based on photometric redshift estimates, although quasars with spectroscopically-determined $z<1$ are included in the final sample.  In addition, the morphological cut imposed on their sample requires quasars to appear stellar in the UKIDSS and SDSS images.  The choice in \citet{Maddox12}  to exclude extended sources is intended to avoid host galaxy contamination which affects quasar colors.  However, \citet[\S2]{Glikman12} showed that the near-infrared-to-optical colors of reddened quasars are largely unaffected by the presence host galaxy light, because the longer wavelengths are still dominated by the quasar continuum.  

Of the 14 quasars found in this work, only UKFS0016$-$0038 is classified as stellar by both UKIDSS and SDSS.  Only two UKFS quasars are classified as having stellar morphology ({\tt mergedClass = -1}) in UKIDSS and another two quasars are classified as probableStar ({\tt mergedClass = -2}).  The remaining eight quasars are classified as galaxies and would have been excluded by \citet{Maddox12} and \citet{Banerji12} (including UKFS0158$+$0027 at $z=1.35$).  As we noted in \citet{Glikman12}, imposing morphological criteria on red quasar candidates selection schemes may exclude a large fraction of the sources, particularly those at low redshifts or in a post-merger phase.  

\subsection{Mid-Infrared Selection}\label{sec:wise}

\begin{figure}
\epsscale{1.2}
\plotone{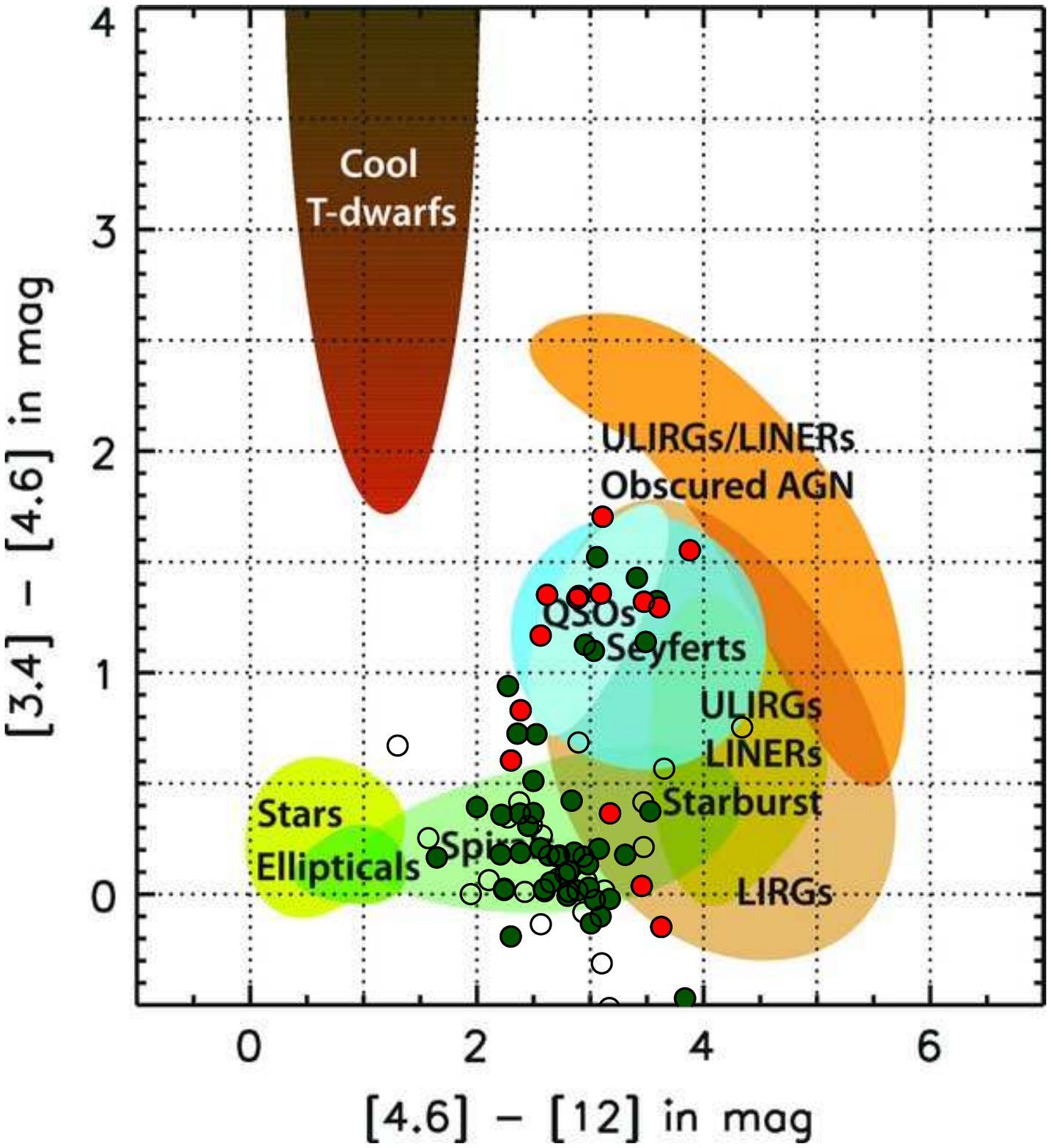}
\caption{Infrared WISE color-color space from Figure 12 of \citet{Wright10} showing the locations of various classes of astrophysical objects with UKFS sources over-plotted circles.  Confirmed red quasars  are shown in red and lie more or less where all quasars are expected to be found, apart from a few outliers (see text).  Spectroscopically-observed objects that do not show broad emission lines are colored green, while unobserved candidates are plotted with open circles.}\label{fig:wise}
\end{figure}

An alternative method for finding quasars unbiased by dust extinction is to use their mid-infrared colors.  Work by \citet{Lacy04} and \citet{Stern05} using the {\em Spitzer} Space Observatory revealed that the power-law nature of quasar spectra can be exploited toward even longer wavelengths.  Both of these studies in near-to-mid infrared color space find quasars independently of reddening and have been effective at identifying heavily obscured AGN, e.g., Type II sources that do not reveal broad emission lines in their spectra.  Recently \citet{Donley12} improved upon this method to identify large numbers of obscured quasars.  While successful at identifying populations of obscured AGN at high luminosities, they are less effective in deep fields \citep{Cardamone08}.  The small areal coverage of {\em Spitzer} surveys were therefore unable to identify the rare and luminous quasars found by, e.g., the F2M survey.  The recent all-sky data release from the Wide-Field Infrared Survey Explorer \citep[WISE;][]{Wright10} now offers an opportunity to identify rare luminous systems of the kind we have found here in the UKFS sample.  

To examine the colors of the UKFS quasars in the mid-infrared, we matched the UKFS sample (87 objects) to the WISE all-sky catalog; 84 have matches within 2\arcsec, including all fourteen confirmed quasars.  Figure \ref{fig:wise} shows the location of the UKFS candidates plotted in the WISE W1$-$W2 vs.\ W2$-$W3 color-color space (corresponding to $[3.4\um] - [4.6\um]$ vs.\ $[4.6 \um] - [12 \um]$ bands). \citet{Wright10} showed that this color-space is effective at separating extragalactic sources from stars and brown dwarfs.  

The UKFS red quasars are plotted with red circles and lie mostly in the area where quasars are expected to be found.  This means that the WISE color selection can be effective at finding quasars independent of reddening. However, some quasars fall outside the quasar space.  Two are bluer in both W1$-$W2 and W2$-$W3, and another three lie in the overlapping region between spirals and Luminous Infrared Galaxies (LIRGs). 
 As the postage-stamp images of the UKFS quasars presented in Figure \ref{fig:sed1} show, some of the UKFS quasars have close companions, possibly indicative mergers giving rise to the reddening and triggering of these quasars.  Since the WISE point-spread-function is $\sim 6$\arcsec\ in W1, W2 and W3, the light from any close companions is blended with the quasar light and likely affects some objects' infrared colors.  It is curious, however, that the three quasars in the LIRG/spiral region are all at $z\sim 0.6$\footnote{Although the redshifts for these three sources are based on a single broad emission line, H$\alpha$ is the most plausible line given the absence of an optical spectrum or strong lines in the remaining parts of the infrared spectrum}.  In Section \ref{sec:ebv} we consider the effect of significant host galaxy light affecting these sources' SEDs.

UKFS sources with spectroscopic identifications that do not show broad lines are plotted with green circles.  All the objects in the WISE quasar space \citep[W1$-$W2 $>0.8$; c.f.,][]{Assef12} have spectra.  There are six sources in this space whose spectra do not show board line emission, suggesting that while the WISE color selection is effective at finding quasars, it may suffer from some contamination.  Sources with no spectroscopic observation are plotted with open circles. Two red quasars have $0.6 <$ W1$-$W2 $< 0.8$ and $2 < $ W2 $-$ W3$< 3$; one additional object with similar colors in this space has no spectroscopic identification. 
Another four unidentified objects in the ULIRG/LIRG/spiral overlapping region may also be quasars.
The remaining unidentified sources with W1$-$W2$<0.5$ and W2$-$W3$<3$ are likely not quasars.  

\begin{figure}
\epsscale{1.2}
\plotone{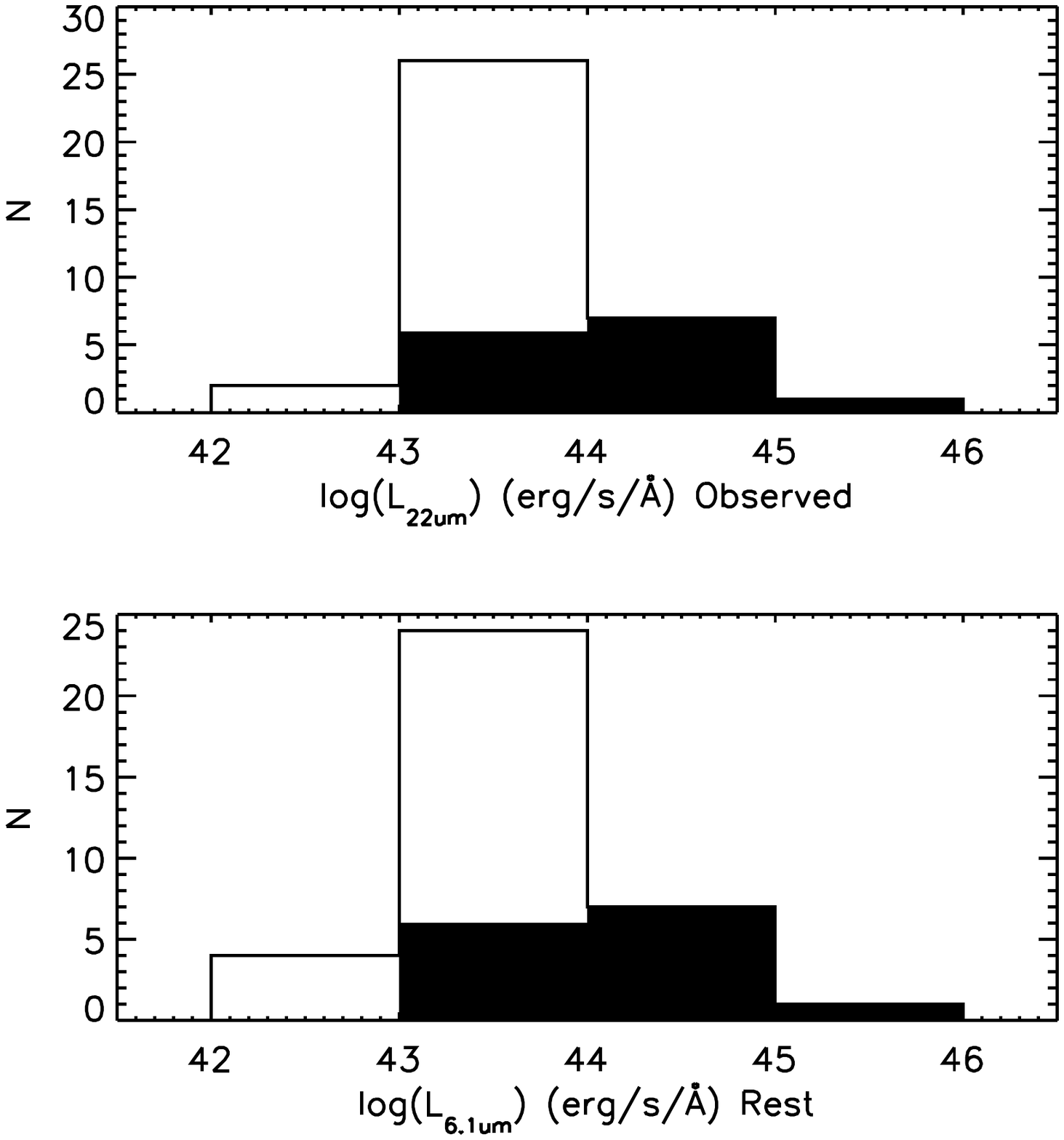}
\caption{Infrared luminosities based on WISE photometry for all UKFS source that have redshifts.  {\em Upper panel:}  shows a histogram of the observed frame 22$\mu$m luminosity.  {\em Bottom panel:} shows the rest frame luminosity at a rest frame of 6.1$\mu$m which is the longest rest-frame wavelength common to all sources.  In both panels, all confirmed quasars (filled histogram) are the most luminous sources. }\label{fig:irlumhist}
\end{figure}

Might some of the objects whose spectra do not show broad lines have infrared luminosities that would place them in the quasar regime?  Such objects would be heavily obscured (e.g., Type II or Compton thick) quasars.  Figure \ref{fig:irlumhist} shows the WISE-derived infrared luminosities of all sources for which we were able to determine a redshift.  Confirmed quasars are overplotted in the filled histogram.  The top panel shows the observed-frame W4 22$\mu$m luminosity.  The bottom panel shows the rest-frame luminosity at 6.1$\mu$m (corresponding to 22$\mu$m at $z=2.5$, the highest redshift in our sample) which we determine by interpolating between the WISE bands.  In both panels, all the quasars are the most luminous sources. Since there are no high luminosity sources with no broad lines  we are confident that we haven't missed any quasars among our candidates.  However, if these sources are obscured by large amounts of reddening (e.g., $E(B-V) \gtrsim 5$, as is typical for Seyfert 2 galaxies) then our estimate of the luminosity at $6.1\mu$m may still be underestimated.  At this wavelength a reddening of $E(B-V)=5$ removes $\sim 40\%$ of the light.  And without knowing the amount of extinction, even WISE does not reach long enough rest frame wavelengths to be unaffected by such high extinction.  

\section{Red Quasar Demographics}\label{sec:surfdens}

\begin{figure}
\epsscale{1.2}
\plotone{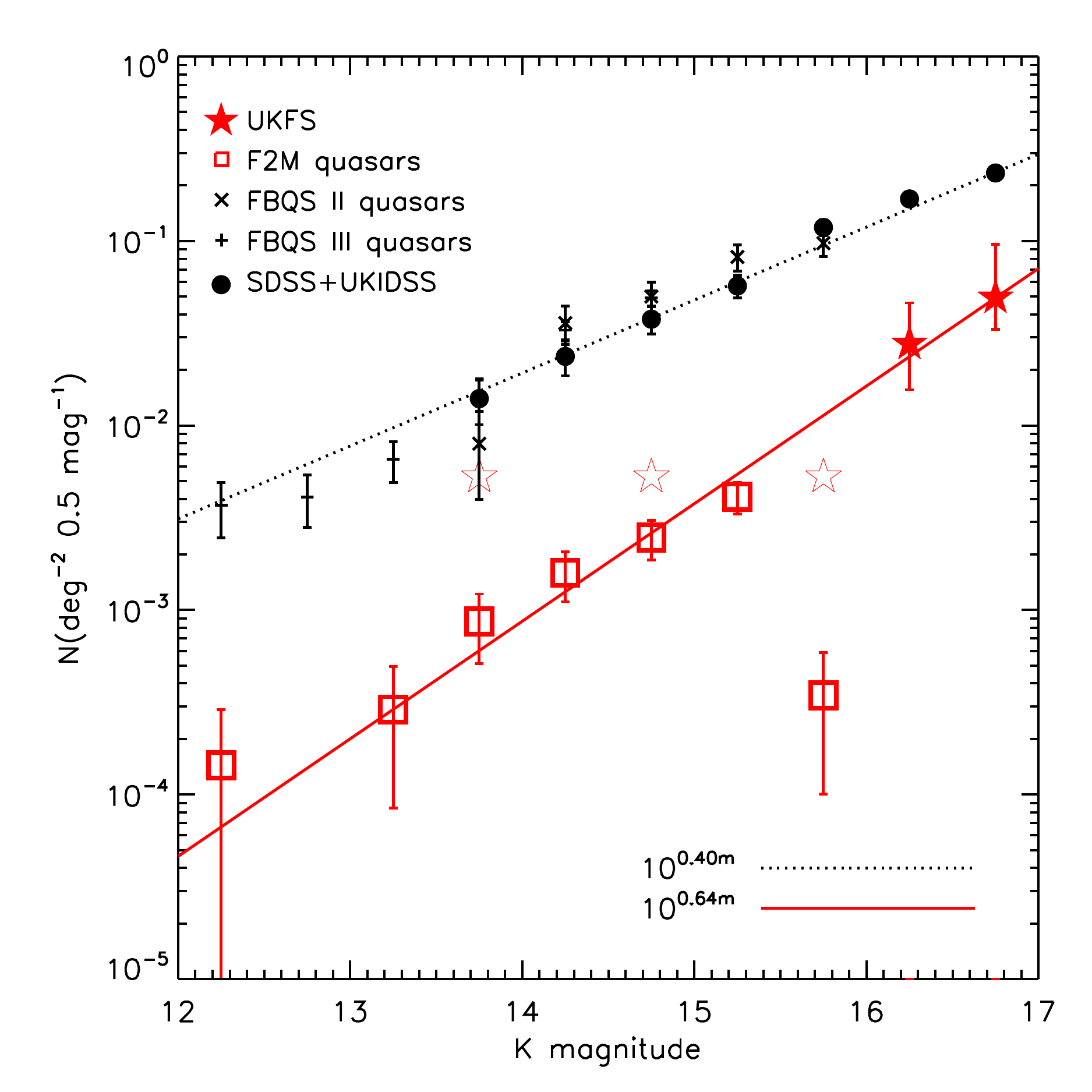}
\caption{Observed surface density of red quasars compared to blue quasars.  The UKFS survey (red stars) goes $\sim 1.5$ magnitudes deeper than F2M (open red squares).  The dotted black and solid red lines are the best power law fit to the points between $K=13.5$ and $K=17$.  We see a rise in the number counts of red quasars toward fainter magnitudes, suggesting that reddening is common in moderate luminosity quasars and is not just a phenomenon associated with the most luminous sources. }\label{fig:sd}
\end{figure}

We compute the surface density of the UKFS red quasars as a function of $K$-band magnitude, extending the measurement of the F2M survey by 1.5 magnitudes fainter in the $K$-band.  The space density of red quasars increases sharply toward fainter magnitudes.  We plot in Figure \ref{fig:sd} the number counts of red quasars compared with optically-selected quasar samples.  The open red squares are the F2M quasars as shown in Figure 10 of \citet{Glikman12}.  The stars represent the results from this study.  The three brightest bins each contain only one source, because of the relatively small area covered by the UKIDSS DR1; we mark these with open stars without their error bars to avoid cluttering the plot.  
The two brightest sources are the two recovered F2M quasars,  
The measurement of the space density at $K=15.5-16$ is highly incomplete in both the F2M survey, as the 2MASS limit is reached, and in this work, as the area probed is too small.  The filled stars represent robust measurements of the surface density of radio-selected red quasars in a previously unexplored magnitude regime: their surface density rises smoothly as a power law with increasing magnitude (faintness). We fit a power law to the red quasars' space densities using the robust least-absolute-deviation linear fitting routine, LADFIT, in IDL, which avoids strong outliers, in the magnitude range $13.5 <K\le17$.  We find that the number of red quasars per half magnitude bin increases as $N\sim 10^{0.64m}$.

To determine the size of the red quasar population relative to normal, blue quasars, we need to construct an appropriate comparison sample.  In \citet{Glikman04,Glikman07,Glikman12} we relied on the FIRST Bright Quasar Survey \citep[FBQS;][]{Gregg96,White00,Becker01} which is a quasar sample constructed from a radio plus optical selection, imposing a $B-R<2.0$ color cut.  That sample was then matched to 2MASS to construct a $K$-magnitude distribution with the same flux limits in the radio and near-infrared as the F2M quasars.  For comparison, we plot the FBQS surface density for FBQS based on their 2MASS magnitudes in Figure \ref{fig:sd}, as in Figure 10 of Glikman et al. (2012).

Since the UKFS sample relies on FIRST plus UKIDSS we must construct an equivalent optically-selected quasar sample.  We utilized the SDSS-UKIDSS-matched quasar catalog of \citet{Peth11}.  This sample contains 20,991 spectroscopically confirmed quasars from the SDSS DR5 quasar catalog \citep{Schneider10} (as well as 130,000 quasar candidates selected using the nine colors from SDSS and UKIDSS) that have photometry in the UKIDSS DR3 LAS catalog over 1200 deg$^2$.  We then matched this catalog to the 16 February 2012 FIRST catalog, which covers 10,635 deg$^2$ and overlaps the \citet{Peth11} quasar catalog completely, to produce a catalog of 1,197 optically-selected, UKIDSS- and FIRST-detected quasars.  We plot the number counts for this comparison sample with filled black circles line in Figure \ref{fig:sd}, scaling the survey area by the 77.4\% spectroscopic completeness of the SDSS quasar survey \citep{Richards06} for an effective area of 928 deg$^2$.  The space density of these quasars is consistent with the FBQS number counts in the areas where the two surveys overlap, providing confidence that we have extracted a comparable, but deeper, sample of blue quasars. The best fit a power-law to the blue quasar number counts is $N\sim 10^{0.40m}$.

To determine the fraction of red quasars out to $K=17$ we integrate the power-law curves for the blue and red quasars out to $K=17$ and find that the space density of red quasars is 0.14 deg$^{-2}$, while for blue quasars it is 0.80 deg$^2$ (considering only FIRST-detected sources).  This result means that the surface density of red quasars is $17\%$ as high as the surface density of blue quasars.  This is a higher fraction than the result from the F2M quasars, which made up $10\%$ of the surface density to $K=14.5$, suggesting that the fraction of red quasars may be increasing with decreasing luminosity and/or redshift.  

As we noted in Section \ref{sec:wise} three of the sources in our sample reveal broad emission at $\lambda \sim 1.08 \mu$m, which we interpret to be H$\alpha$ placing them at $z\sim 0.65$, whose WISE infrared colors are in the overlapping LIRG/spiral region.  We examine the effect of excluding these sources from the surface density analysis, in case our identifications turn out to be incorrect (e.g., once an optical spectrum is obtained and examined) and find that the fraction of red quasars drops to 9\% (more in line with the F2M fraction) and the shape of the power-law fit changes to $N\sim 10^{0.50m}$, which is still steeper than for blue quasars. Since the UKFS survey uses a more restrictive $r-K$ color cut and therefore might be missing some quasars, these surface densities are likely a lower limit and the rise in space density of red quasars toward fainter magnitudes is potentially even higher.  

\section{Reddening and Spectral Energy Distributions}\label{sec:ebv}

\subsection{Fitting a Reddened Quasar Template to the Spectra} 

\citet{Glikman12} explored the effectiveness of several dust laws at fitting the spectral shape of red quasars.  Compared with the Large Magellanic Cloud (LMC) dust law of \citet{Misselt99}, the milky way dust law of \citet{Cardelli89} and the starburst extinction law of \citet{Calzetti94}, the best fit was produced by Small Magellanic Cloud (SMC) reddening law of \citet{Gordon98}.  A similar conclusion was reached by \citet{Hopkins04} who noted that reddened quasar spectra lack the 2175\AA\ silicate absorption feature present in the Galactic and LMC reddening laws.   

In \citet{Glikman12} we measured the dust reddening in the F2M red quasars by fitting a quasar template that was reddened by the SMC reddening law of \citet{Gordon98} to our quasar spectra.   We fit our spectra with a reddened quasar template in the same fashion as in \citet{Glikman12}.  The resultant reddened fits are shown in Figure \ref{fig:spec1}.  We see that, in general, the reddened template traces the shape of the continuum.  Two significant outliers are UKFS0156$-$0058 and UKFS0135$-$0043, which we discuss below.  In four cases we have both an optical and a near-infrared spectrum for the source.  For these objects we determined $E(B-V)$ from the, optical, near-infrared and combined spectra.  Table 2 lists the resultant values of $E(B-V)$ found via all the methods that we use. 

When available, the optical spectrum provides a stronger constraint on the reddening than the near-infrared spectrum, since shorter wavelength light is more sensitive to dust extinction.
However, in this work most of the spectra of UKFS quasars are in the near-infrared, which imposes a weaker constraint on the reddening than if optical spectra were also available for these sources.  For example, two of our quasars' near-infrared spectra are best fit with a {\em negative} $E(B-V)$ despite having clearly red colors (e.g., UKFS0135$-$0043). 
To remedy this issue, we utilize the extensive broad band photometry from SDSS, UKIDSS and WISE, which include thirteen photometric measurements (SDSS $u$, $g$, $r$, $i$, $z$; UKIDSS $Y$, $J$, $H$, $K_s$ and WISE W1, W2, W3, W4) spanning $0.3 - 22 \mu$m.

To compute the reddening for our quasars from their broad band photometric SEDs, we used the five-band SDSS {\tt modelMag} magnitudes and their errors, together with the UKIDSS DR1 LAS 3\arcsec\ aperture magnitudes and their errors plus the WISE All-Sky Source Catalog photometry with photometric errors.  We shifted the UKIDSS magnitudes to the AB system by their zero-point offsets provided in Table 7 of \citet{Hewett06} and Table 1 of \citet{Jarrett11} to be consistent with SDSS, and computed their flux densities, $F_\lambda$, in units of erg s$^{-1}$ cm$^{-2}$ \AA$^{-1}$ using the equation:

\begin{equation}
F_\lambda = 10^{[-0.4(m_{AB} + 2.406 + 5\log\lambda_{AB})]},
\end{equation}
where $m_{AB}$ is the quasar's apparent magnitude and $\lambda_{AB}$ is the effective wavelength of the given magnitude's bandpass.  We used $\lambda_{AB} = 3551$\AA, $4686$\AA, $6165$\AA, $7481$\AA, $8931$\AA\ for the SDSS $u$, $g$, $r$, $i$, $z$ bands, respectively,  $\lambda_{AB} = 10305$\AA, $12483$\AA, $16313$\AA, $22010$\AA\ for the UKIDSS $Y$, $J$, $H$, and $K$ bands, respectively \citep{Hewett06} $\lambda_{AB} = 33526$\AA, $46028$\AA, $115608$\AA, $220883$\AA, for the WISE W1, W2, W3 and W4 bands, respectively \citep{Wright10}. 

To compute the corresponding unreddened template colors, we passed the quasar spectral energy distribution (SED) from \citet{Richards06} through the transmission curves for each of the nine bandpasses in the quasar's observed frame, in order to obtain a synthetic magnitude of a standard blue quasar at the same observed wavelengths of the UKFS quasars.   Converting these magnitudes to fluxes resulted in a thirteen-point SED which was used to compare with our observed SEDs.  We use these SEDs to determine the reddening in each quasar in the same manner as with the spectra, except with fewer data points (and weighting by their photometric errors expressed as flux errors). 

We plot the rest-frame optical-through-infrared SED of the fourteen UKFS quasars with black circles with error bars in Figure \ref{fig:sed1}.  The blue curve is the mean SED for all quasars in the \citet{Richards06} sample\footnote{The choice of SED \citep[there are six sub-grouped SEDs in addition to the total SED in ][]{Richards06} does not impact our measurement of $E(B-V)$ significantly.  We tested this by using the different templates to determine $E(B-V)$; all the measured values agree within 0.03 magnitudes.}, with the synthetic photometric measurements plotted with blue squares.  We use the ratio of the thirteen point flux arrays from the measured data and the SED to determine the reddening,
\begin{equation}
E(B-V) = -\frac{1.086}{k(\lambda)} \log\Bigg[\frac{f(\lambda)}{f_0(\lambda)}\Bigg], \label{eqn:ebv}
\end{equation}
where $k(\lambda)$ is the SMC dust law.  We plot the reddened SED in red.  The figure inset is the RGB cutout of size $11.9\arcsec \times 11.9\arcsec$ obtained from SDSS and displayed with an inverted grayscale; several of the images show evidence for interaction.
 
 \begin{figure*}
\plotone{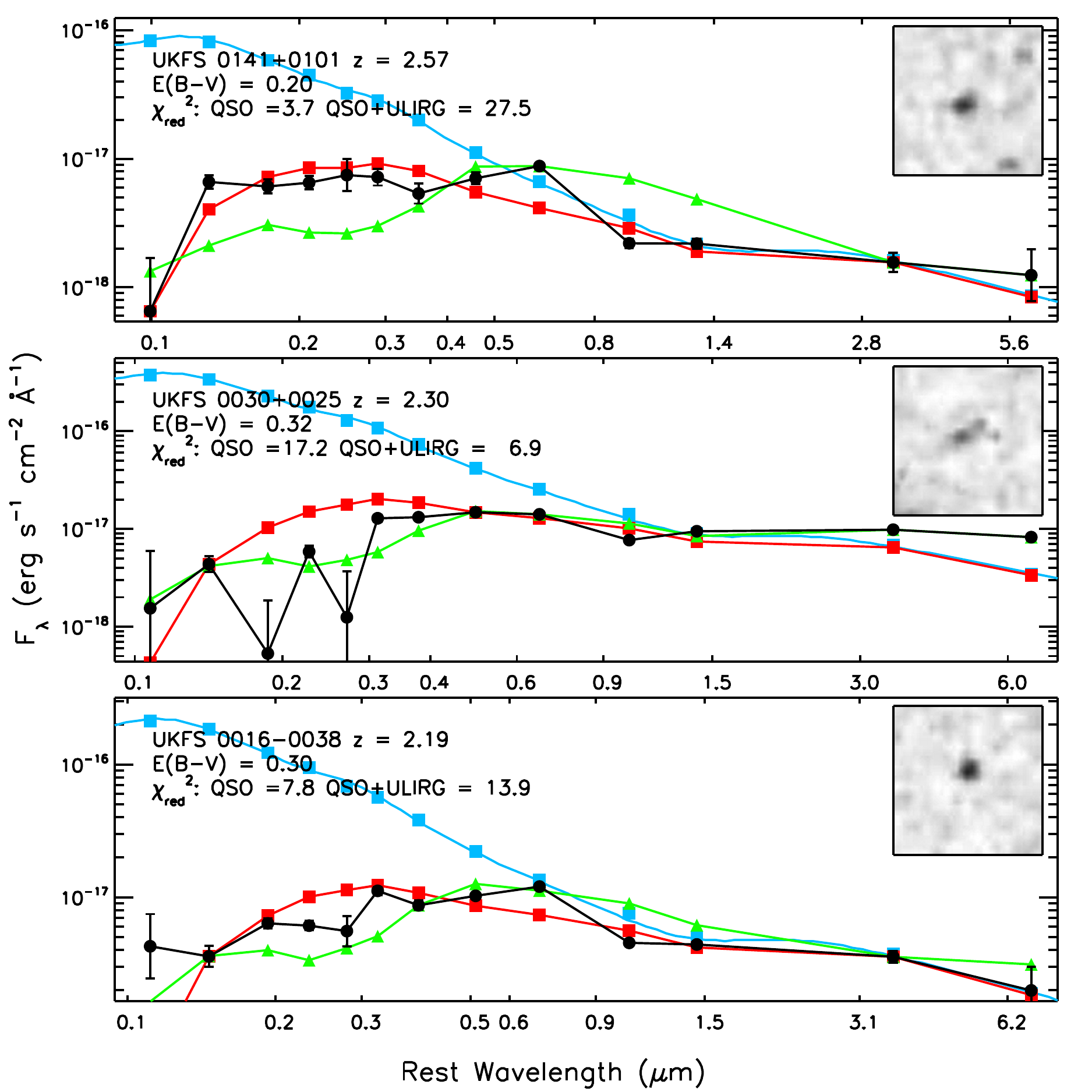}
\caption{Rest-frame optical-through-infrared SEDs for UKFS red quasars (black lines) determined from SDSS ($u$,$g$,$r$,$i$,$z$), UKIDSS ($Y$,$J$,$H$,$K$) and WISE (W1 [3.4$\mu$m],W2 [4.6 $\mu$m],W3 [12 $\mu$m],W4 [22 $\mu$m]) photometry, and ordered by decreasing redshift.  The blue line is the quasar SED from \citet{Richards06}, over plotted with synthetic SDSS, UKIDSS and WISE photometry (blue squares).  The red line shows the best-fit reddened quasar template to the photometric points.  The green line shows the best two-component fit including a reddened quasar plus the starburst/ULIRG galaxy template from \citet{Polletta07}. 
The image inset is an inverted 11.9\arcsec $\times$ 11.9\arcsec grayscale image from SDSS.}\label{fig:sed1}
\end{figure*}

\begin{figure*}
\figurenum{8b}
\plotone{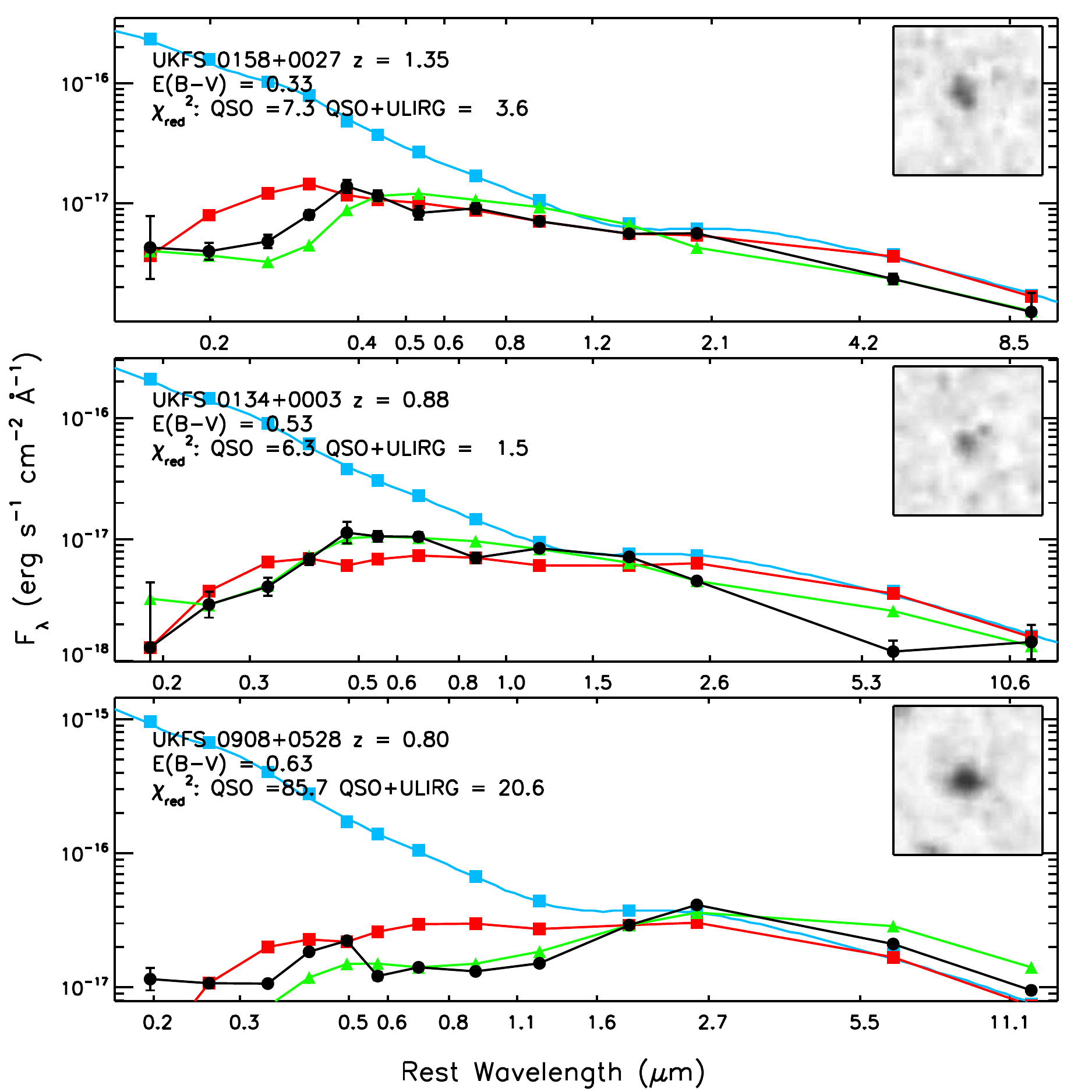}
\caption{{\em Continued.} Optical-through-near-infrared spectra of UKFS quasars. }\label{fig:sed2}
\end{figure*}

\begin{figure*}
\figurenum{8c}
\plotone{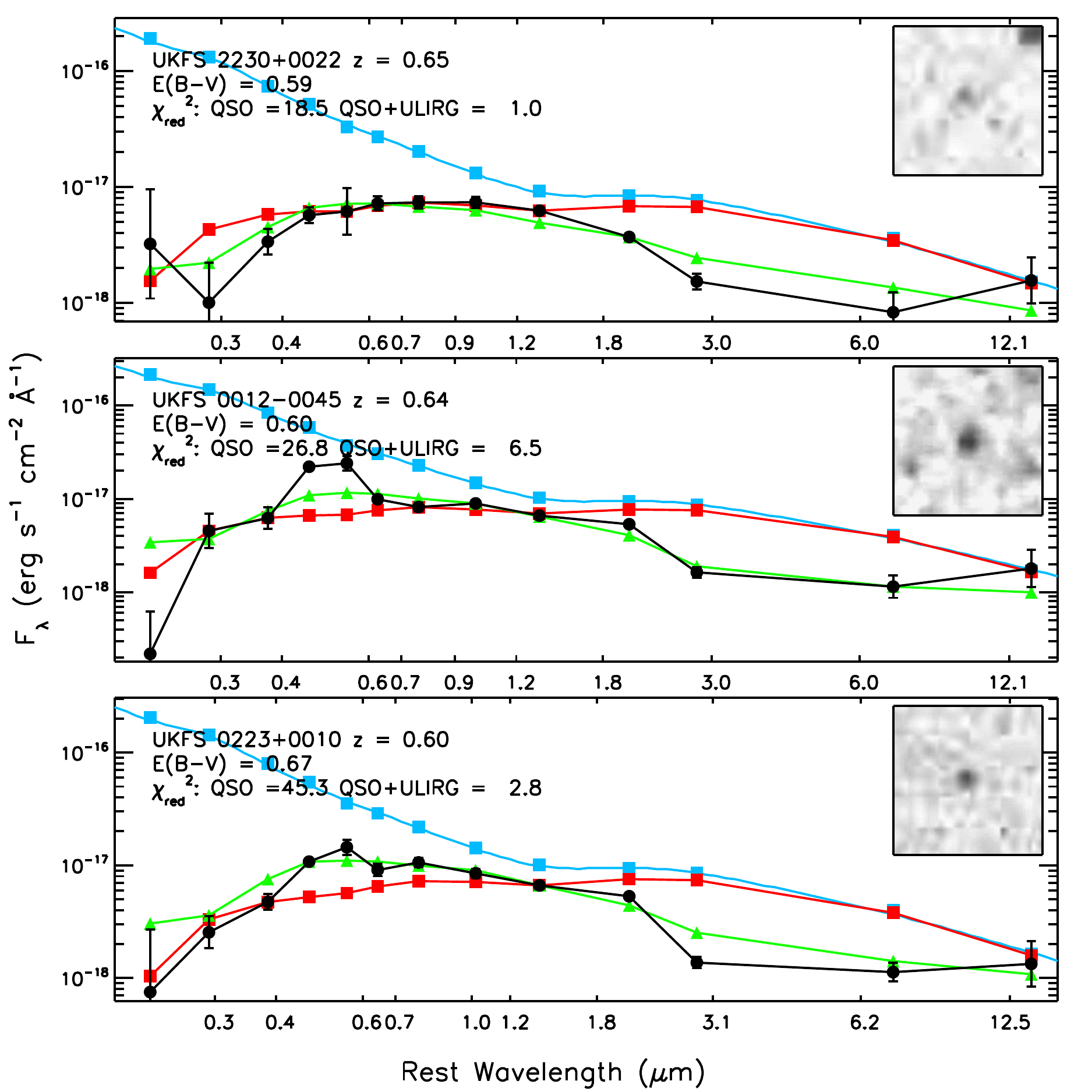}
\caption{{\em Continued.} Optical-through-near-infrared spectra of UKFS quasars. }\label{fig:sed3}
\end{figure*}

\begin{figure*}
\figurenum{8d}
\plotone{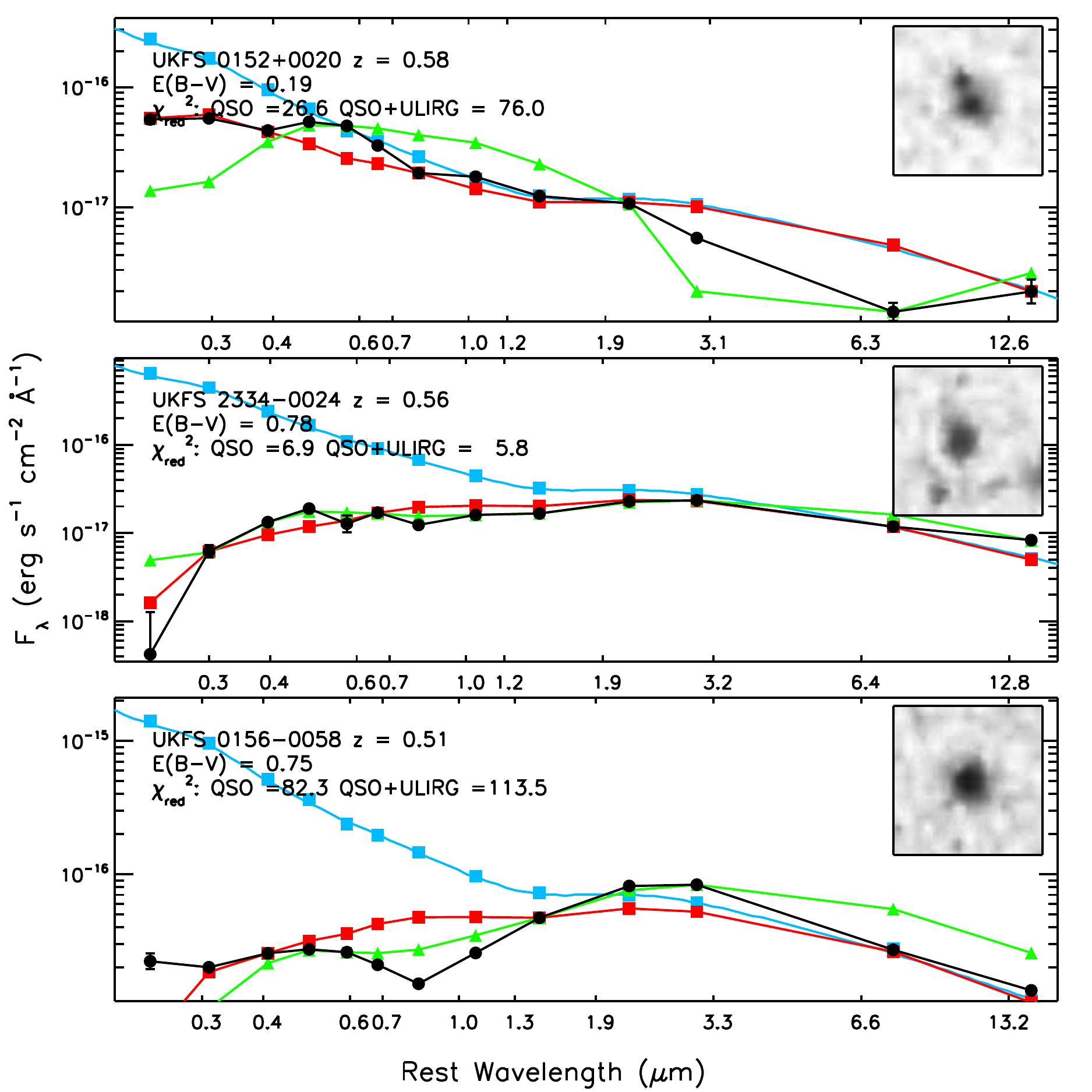}
\caption{{\em Continued.} Optical-through-near-infrared spectra of UKFS quasars. }\label{fig:sed4}
\end{figure*}

\begin{figure*}
\figurenum{8e}
\plotone{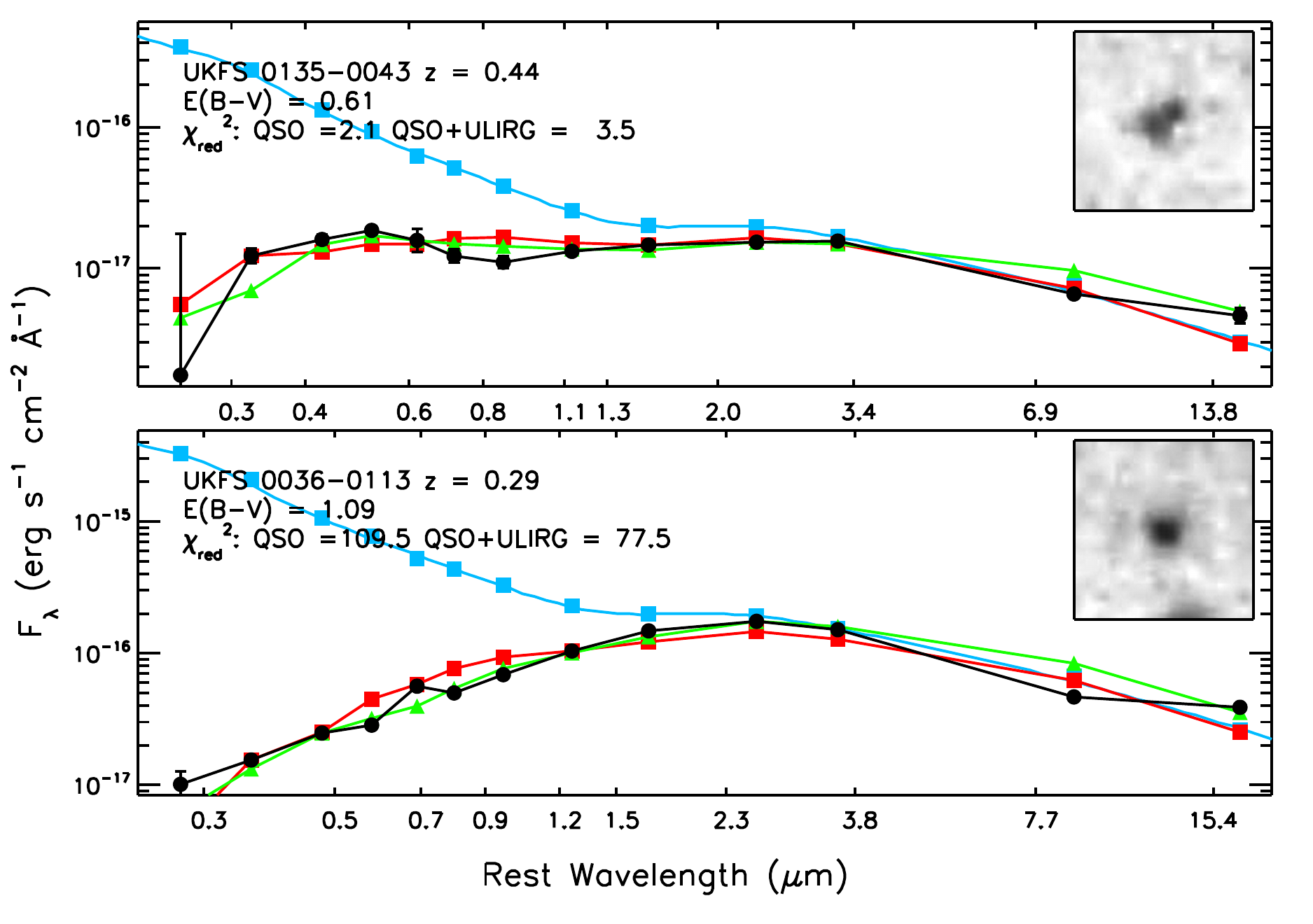}
\caption{{\em Continued.} Optical-through-near-infrared spectra of UKFS quasars. }\label{fig:sed5}
\end{figure*}

The two sources, UKFS$0156-0058$ and UKFS$0135-0043$ have highly divergent spectral fits (Figure \ref{fig:spec3} and Figures \ref{fig:sed4} and \ref{fig:sed5}).  UKFS$0156-0058$ shows a strong red continuum in its near infrared spectrum.  However, this rise does not continue toward the mid-infrared.  Rather, the SED shows a bump with a peak around $\sim 2.5 \mu$m.  If this feature is a signature of hot dust close to the sublimation radius, then its temperature would be $\sim 1200$ K, which has been observed in luminous quasars \citep{Glikman06,Netzer07,Mor11} as well as F2M red quasars \citep{Urrutia12}, but not usually at such high luminosities so as to surpass the direct emission from the AGN itself.  In the case of UKFS$0135-0043$, the SDSS image shows a close double source with a red and blue component.  These sources are so close that they fall within the SDSS fiber diameter of 3\arcsec.  Given the unknown contributions from the blue and red objects -- which may be evidence for a merging system -- it is not surprising that this object does not yield consistent $E(B-V)$ values from the different fits.   We note, however, that the SED fit yields $E(B-V) = 0.61$ while the optical and combined spectral fits yield $E(B-V) = 0.69$ and 0.70, respectively.  Only the $E(B-V)$ from the fit to the near-IR spectrum diverges with a value of $-0.22$.

Using our estimates of the $E(B-V)$ from the SED fits we de-redden the UKFS quasars and examine their intrinsic properties.  
In Figure \ref{fig:abskz} we plot the de-reddened absolute $K$-band magnitudes (in the observed frame) for the UKFS quasars versus redshift.  
The sources are color-coded by the amount of reddening, with yellow circles representing a small amount of reddening ($E(B-V)\sim 0.1-0.5$) red circles representing large amounts of reddening ($E(B-V)\sim 1-1.5$) and orange in between.   
This Figure plots the F2M quasars \citep[similar to Figure 15 of][]{Glikman12}, and the UKFS quasars emphasized with thick black circles.   The small black points represent the FBQS and SDSS comparison samples described in Section \ref{sec:surfdens}.  The dotted lines shows our sensitivity to finding quasars with different amounts of reddening down to the survey limit of $K=17$.  The dashed line shows the 2MASS survey limit of $K=15.5$.  

To evaluate the goodness of the SED fits, we compute the reduced $\chi^2$ statistic (dividing $\chi^2$ by eleven agrees of freedom:  thirteen photometric points, minus the normalization and $E(B-V)$, which are the two free parameters of our fits).  
We quote this value in the legend of each panel ($\chi^2_{\rm red}:$ QSO).  
We see that in general, the reddened quasar SED (red line in Figure \ref{fig:sed1}) does not fit well the shape of the measured SED for the UKFS quasars (black line) across the full wavelength range.  If these systems are indeed quasars in a post-merger early evolutionary phase, then their SEDs -- especially in the infrared -- may be strongly affected by star formation signatures, e.g., hot dust and PAH emission.  Detailed infrared spectroscopy of Palomar-Green quasars  \citep{Schweitzer06,Netzer07} as well as a small sample of F2M red quasars \citep{Urrutia12} find that there is great diversity among the mix of star formation and AGN contributors, although the AGN dominates in most of the F2M red quasars  (with $L_{\rm QSO}/L_{\rm FIR~SB} >2 - 60$ for all but one source).  This becomes more complicated with lower luminosity quasars whose relative contributions from stars and nuclear activity begin to rival each other.  

\subsection{The Effect of a Host Galaxy Component to the Reddening Fits}

To investigate the impact of a host galaxy on our fits, we followed the approach in \citet[Eqn. 3]{Glikman07} and tried adding emission from a host galaxy to the model: 
\begin{equation}
f(\lambda) = Af_{\rm gal}(\lambda) + Bf_{\rm QSO}(\lambda)e^{-\tau_\lambda} \label{eqn:qso_gal}.
\end{equation}
We used the the galaxy templates from \citet{Polletta07} and settled on the Starburst/ULIRG Arp220, whose deep silicate absorption and hot dust bump beyond 10 $\mu$m resembles the features in the WISE photometry seen in some of our sources.
 For the three $z\sim 0.6$ sources, UKFS2230+0022, UKFS0012$-$0045, and UKFS0223+0010, that lie in the LIRG region of Figure \ref{fig:wise}, the fit improved significantly with the addition of a ULIRG template, primarily because of a dip at WISE wavelengths that could correspond to a silicate absorption feature commonly seen in ULIRGs.   Their implied galaxy luminosities, while high, are within plausible limits for LIRGs ($M_K = M_{H \rm rest} =  \simeq -25.9$ or $L_{\rm near-IR} \simeq 5\times10^{11} L_\odot$).  We plot the quasar component of the two-component fit for these sources with colored squares in Figure \ref{fig:abskz}\footnote{We plot these sources with circles and squares to represent their derived values from both models.} and add a row in Table 2 listing their reddening and luminosities from the two-component fit. They are significantly redder ($E(B-V) \simeq 1 - 2$) and less luminous ($M_K \simeq -24.9$ to $-25.6$).
 
In another three quasars, UKFS0158+0027, UKFS0134+0003, UKFS0908+0528, the more complex model yields a lower $\chi^2$ (labeled QSO+ULIRG in Figure \ref{fig:sed1}).  We also plot these sources with squares in Figure \ref{fig:abskz} and list their reddening and luminosity values in Table 2. In UKFS0158+0027 at $z=1.350$ (shown in the top panel of Figure \ref{fig:sed2}), the two-component model (green line) yields a better fit than the quasar-only model.  However, the reddening determined from this is $E(B-V) = 3.90$, which implies a de-reddened quasar luminosity of $M_K = -31.94$.  Such a high reddening value is unlikely given that we see strong, broad ($\sim 2000$ km s$^{-1}$) H$\alpha$ line in its near-infrared spectrum.  For UKFS0135+0003, shown in the middle panel of Figure \ref{fig:sed2}, the inclusion of a host component to the fit also improves $\chi^2$ significantly.  In addition, the reddening, $E(B-V) = 1.17$, and de-reddened quasar luminosity, $M_K = -26.84$, are reasonable.  The WISE colors of this source (W1$-$W2$=0.83$ and W2$-$W3$=2.38$) place it just outside the quasar region, on the blue end, not near the LIRGs, but its image shows a nearby companion, likely contaminating the WISE photometry (which is not well fit by either model).  Finally, UKFS0908+0528 does have an improved $\chi^2$ when fit by a two-component model.  But neither model produces a satisfactory fit to the optical photometry.  Furthermore, the reddening derived from the two-component fit is $E(B-V) = 2.91$ with a de-reddened quasar luminosity of $M_K = -29.50$.  This reddening also seems high given the strong, broad ($\sim 2600$ km s$^{-1}$) H$\alpha$ seen in its near-infrared spectrum. 
 
Therefore, we conclude that 10 of 14 quasars are most likely dominated by quasar continuum emission, whereas in the other four (three LIRG-like sources and UKFS0135+0003), the galaxy could contribute significantly.  For the remaining ten sources, the results of the two component fit result in extremely luminous host galaxies ($M_K \simeq -27$ to $-30$).  Furthermore, the two-component fits imply that the quasar is far more heavily reddened than the single component fit, with $E(B-V)\simeq2-6$, which translate into extinction-corrected luminosities that are as high or higher than what we estimate from the single component fit.  In addition, all of our sources exhibit broad emission lines, so we expect a quasar SED to be the dominant component. 

If instead these 10 red quasars actually do lie in galaxies that are more luminous than has ever been seen before (the $z>2$ sources would have $L>10^{13}L_\odot$ in the $R$-band), such an extraordinary claim could be investigated with better sampled photometry. We also note that, when the SED has been better defined, with {\em Spitzer} IRS spectra \citep[Section 4 of][]{Urrutia12}, the galaxy contribution was negligible.  At present, our 13 photometric data points prevent us from disentangling the true nature of the reddened quasar SEDs, which are more complicated than a two component model can describe.  These objects are in the company of other newly-discovered, complicated systems that are not yet fully understood and not well matched to any known SEDs, e.g. the ``hot DOGs" in \citet{Wu12} or the dust rich quasars in \citet{Dai12}.

\citet{Glikman07,Glikman12} identified several quasars whose flux variability between the epochs of their optical and near-infrared observations mimicked reddening, but which, upon spectroscopic observations, revealed a normal blue quasar.  These objects accounted for $<10\%$ of the sample.  Although it is  possible that quasar variability may have contaminated the UKFS candidate sample and possibly affected our reddening estimates, examination of the data suggests that it is not a significant issue.  None of our spectroscopically-confirmed quasars appear to have strong blue continua.  
In addition, UKIDSS observed two filters simultaneously $Y$,$J$ and $H$,$K$ \citep{Dye06}. Therefore, while the $J$ and $K$-band images were taken at different times, we see no sharp discontinuities from $Y$ and $J$ to $H$ and $K$.  Furthermore, the SDSS photometry is taken near-simultaneously as a result of the drift-scanning design of the survey \citep{York00} (WISE bands are also  observed simultaneously).  And since the reddening is most sensitive to the shorter wavelengths in the SED, it is the SDSS photometry that most constrains our estimates of $E(B-V)$, we therefore rule out strong variability effects.

Another, more challenging, explanation for divergences seen between the observed quasar SEDs and the reddened template SEDs in Figure \ref{fig:sed1} is that the dust law is not well known.  We use the SMC dust-law because it is empirically the best fit to red quasar spectra.  However, a better understanding the composition and extinction properties of the dust obscuring red quasars is needed, and beyond the scope of this paper.  For what follows, adopt the (relatively conservative) $E(B-V)$ values from the single component fits to a quasar SED in the discussion that follows.  

\section{Where are the heavily-reddened high-redshift quasars?}

\begin{figure}
\epsscale{1.2}
\plotone{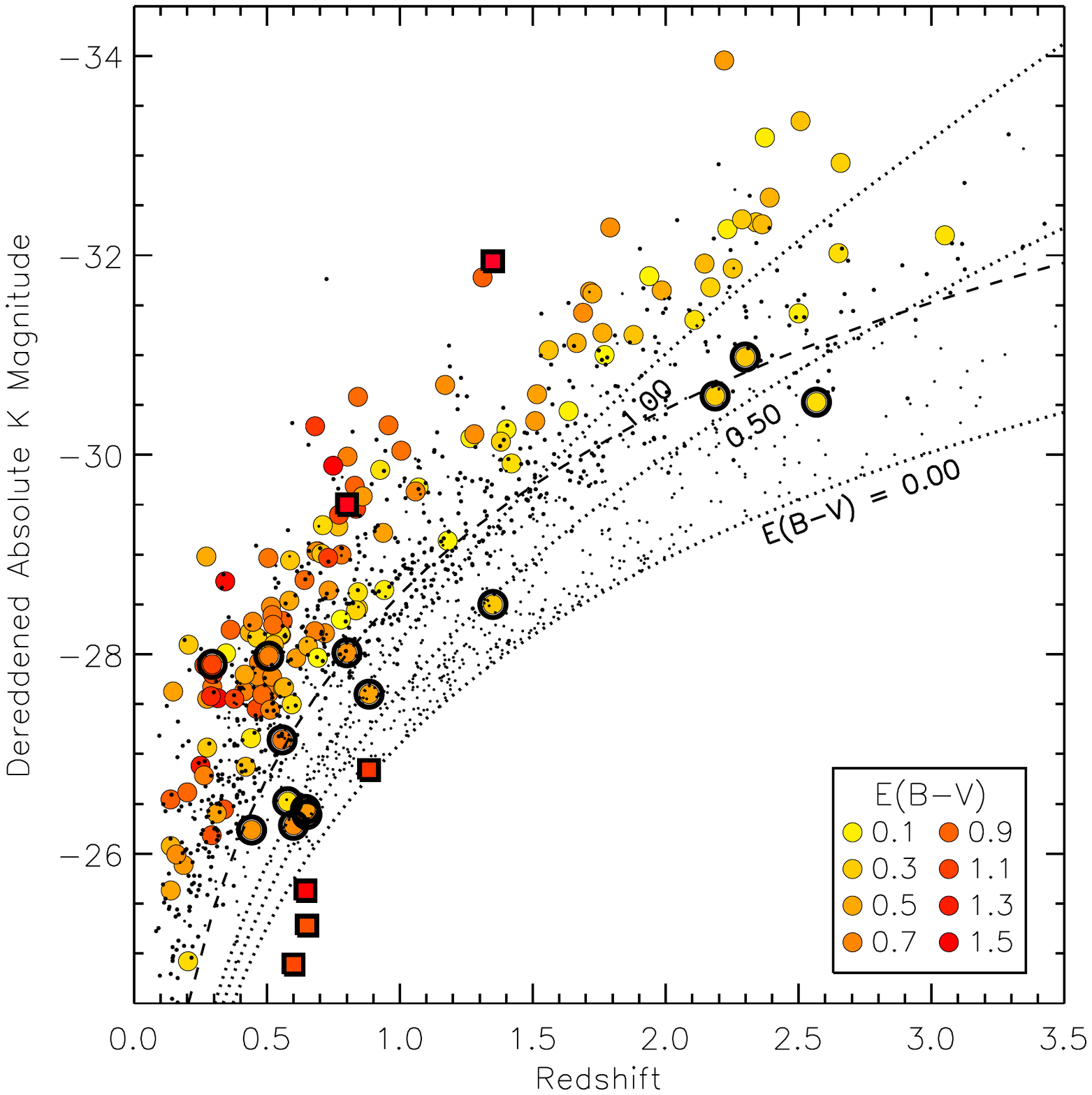}
\caption{De-reddened absolute $K$ magnitude versus redshift for red quasars.  UKFS quasars are identified by thick black circles, the remaining circles are the F2M quasars from Glikman et al. (2012).  The circles are color coded by the amount of reddening, based on the broad band photometric fits and annotated in the legend.  The quasar component of the three sources whose WISE colors are consistent with LIRGs and whose SEDs are better-fit by a two-component model are plotted with square symbols.  The small black points are the same FBQS and SDSS quasars that are plotted in Figure \ref{fig:sd}.  The dashed line indicates the sensitivity limit of the F2M survey ($K=15.5, E(B-V) = 0$) while the dotted lines show the sensitivity limit of UKFS ($K=17$) with increasing amounts of extinction ($E(B-V) = 0.0, 0.5, 1.0$). We recover moderately-reddened quasars at high redshift ($z>2$) that lie just below the 2MASS limit.}\label{fig:abskz}
\end{figure}

Our intention with this red quasar survey was to recover heavily reddened ($E(B-V) \gtrsim 1$) quasars at high redshift ($z\gtrsim 2$) that were not found in the F2M survey.  Figure \ref{fig:abskz} shows that we do find heavily reddened quasars at low redshifts ($z\lesssim 1$), but at $z>2$ our sources are all moderately reddened.  The SED fitting issues that arose for many of the sources at $z<2$ are not a concern for the three highest redshift quasars, as the quasar-only model produces reasonable fits and the two-component fits yield unphysically luminous hosts.  Why don't we find heavily reddened quasars at high redshifts, despite our deeper survey? 

Because of the relatively shallow $K\simeq 15.5$ limit of 2MASS, we had found only lightly reddened quasars at high redshifts.  In addition, these objects need to be at the very luminous end of the quasar luminosity function (QLF), and are therefore extremely rare: there are only 10 F2M quasars with $E(B-V)>0.2$ and $z>2$ over 9030 deg$^2$.  
However, even with this deeper UKIDSS limit, the quasars at $z\gtrsim 2$ with $E(B-V) \gtrsim 1$ still must be extremely luminous ($M_K \gtrsim -31.0$).  Because these objects are very rare -- the space density of FBQS quasars with $1.7 < z < 2.7$ and $M_K<-31$ is $\rho_{\rm QSO} = 0.013$ deg$^{-2}$ -- they are not likely to be found in 190 deg$^2$ of this survey.  Assuming that the same reddening distribution exists at high redshift as we see in the F2M sample at low redshift (e.g., $z<0.8$), and assuming that the space density of red quasars make up $\sim 20\%$ compared to unreddened quasars, then we can estimate the area, $A$, needed for a survey to detect at least one heavily reddened quasar at $z\sim 2$:.
\begin{equation}
A = 1/(0.2 \times \rho_{\rm QSO} \times f_{E(B-V)}), \label{eqn:red_area}
\end{equation}
where $f_E(B-V)$ is the fraction of red quasars redder than a chosen $E(B-V)$ based on the distribution of reddenings of F2M red quasars at low redshift.  Equation \ref{eqn:red_area} shows that one needs at least $\sim 500$ deg$^2$ to find one quasar with $E(B-V)>0.5$, given that $f_{E(B-V)} = 0.68$.  More heavily reddened quasars ($E(B-V)\ge1$) are even rarer, with $f_{E(B-V)} = 0.2$, requiring $\sim 2000$ deg$^2$ of survey area to find a single such source.  

These estimates are conservative, based on reddenings derived from our single component quasar SED fits.  If some host galaxy component is present in the SEDs then the quasar component would be redder (in order to conserve the total observed flux), which means that less area may need to be probed. It is also likely that there is evolution in the reddening distribution and that the density of reddened sources increases with redshift, as the merger rate rises toward $z\sim 2$.  However, a larger sample than the one presented here is needed to disentangle the effects of the $K$-correction, evolution of the QLF, selection effects and small number statistics before any statements about the evolution of reddening can be made.  

Of course, a survey that does not rely on the radio would increase the sample size by a factor of $\sim 10$, having the same effect as increasing the survey area by the same factor (assuming the radio emission is independent of reddening).  Eliminating the radio-detection requirement is not trivial, however, as the number of contaminants rises significantly, especially if the morphological criterion is also removed.  We explored the effect of dropping our FIRST detection criterion in the UKIDSS DR1 by selecting sources with $(J-K)_{\rm Vega} > 1.2$, $g_{\rm AB} - J_{\rm Vega} > 1.9$ and $K_{\rm Vega}<17$ and no morphological restriction to see if we recover any red quasar spectra in public spectroscopic databases.  Out of 6857 sources obeying the aforementioned criteria (which were not visually inspected) only ten had a spectrum in the SDSS DR9 spectroscopic database; three are classified as early type galaxies and the remaining seven are stars (of spectral types K5 through M9).  Another three emission-line galaxy spectra were identified from the WiggleZ Dark Energy Spectroscopic survey \citep{Drinkwater10} as well as 136 LRGs in the 2SLAQ catalog \citep{Cannon06}.  No quasars were found.   We conclude that radio selection affords us great efficiency and additional wavelength constraints, such as infrared color selection with WISE, would be required in order to effectively find radio quiet red quasars. 

\section{Conclusions}

We have presented a pilot survey -- the UKFS survey -- to identify reddened quasars in the FIRST and UKIDSS survey, initially focusing on the 190 deg$^2$ of the UKIDSS first data release.  Combining these data with optical photometry from SDSS, we applied the color cuts $r-K>5$ and $J-K>1.5$ and selected 87 candidates with $K\le17$.  We have spectroscopic observations of 64 of our candidates, amounting to 74\% completeness, but we are 95\% spectroscopically complete below $K=16.5$ mags.  The UKFS survey finds 14 quasars, eight of which are presented here for the first time.  Their redshifts extend to $z\sim 2.5$ and their space density rises steeply toward fainter magnitudes.  We find that red quasars make up 17\% of quasars based on their {\em apparent} magnitudes. If we exclude the three LIRG-like sources whose nature is more ambiguous, the space density falls to 9\%.  Our sample is not large enough to extract the extinction-corrected $K$-band distribution or the intrinsic fraction of red quasars.  

We compare our method of red quasar selection to the KX-method and find that the methods are consistent.  However, any red quasar selection technique that restricts candidates to morphologically stellar sources will miss most red quasars.  Most of the quasars in our sample are not classified as stellar in SDSS or UKIDSS.  However, including candidates with extended morphologies adds significant numbers of red galaxy contaminants to the survey.

We examine the infrared colors of the red UKFS quasars from WISE and find that their W2$-$W1 vs.\ W1$-$W2 colors are mostly consistent with those of unreddened quasars, though some sources have colors more similar to LIRGs and/or spiral galaxies.  Combining WISE colors with optical to near-infrared color selection  minimizes contamination from red galaxies yet allows us to still include extended morphologies and to drop the requirement of a radio detection.  

We analyze the SEDs of the UKFS quasars and use broad band photometry along with optical and near-infrared spectroscopy to derive their reddening, $E(B-V)$.  Even with the increased depth of UKIDSS we do not find heavily reddened ($E(B-V)\gtrsim0.5$) quasars at high redshifts ($z>2$).  To find the heavily reddened quasars at high redshifts, we require either a larger area survey, a deeper flux limit, and/or a longer wavelength selection that is less affected by dust.  The results of this survey is a first step toward this end.

\acknowledgments

We thank Nadia Lara whose summer research project -- supported through the Caltech's FSRI program -- helped with the candidate selection.  
EG is supported by an NSF Astronomy and Astrophysics Postdoctoral Fellowship under award AST-0901994.
SGD acknowledges a partial support from the NSF grant AST-0909182.
We are grateful to the Palomar Observatory staff for their assistance during observing runs.  We thank Yale University for support of SDSS-III participation. \\

The National Radio Astronomy Observatory is a facility of the National Science Foundation operated under cooperative agreement by Associated Universities, Inc.

The UKIDSS project is defined in \citet{Lawrence07}. UKIDSS uses the UKIRT Wide Field Camera \citep[WFCAM][]{Casali07}. The photometric system is described in \citet{Hewett06}, and the calibration is described in \citet{Hodgkin09}. The pipeline processing and science archive are described in Irwin et al (2009, in prep) and \citet{Hambly08}. 

Funding for the creation and distribution of the SDSS Archive has been provided by the Alfred P. Sloan Foundation, the Participating Institutions, the National Aeronautics and Space Administration, the National Science Foundation, the U.S. Department of Energy, the Japanese Monbukagakusho, and the Max Planck Society. The SDSS Web site is http://www.sdss.org/.

The SDSS is managed by the Astrophysical Research Consortium (ARC) for the Participating Institutions. The Participating Institutions are The University of Chicago, Fermilab, the Institute for Advanced Study, the Japan Participation Group, The Johns Hopkins University, the Korean Scientist Group, Los Alamos National Laboratory, the Max-Planck-Institute for Astronomy (MPIA), the Max-Planck-Institute for Astrophysics (MPA), New Mexico State University, University of Pittsburgh, University of Portsmouth, Princeton University, the United States Naval Observatory, and the University of Washington.

Funding for SDSS-III has been provided by the Alfred P. Sloan Foundation, the Participating Institutions, the National Science Foundation, and the U.S. Department of Energy Office of Science. The SDSS-III web site is http://www.sdss3.org/.

SDSS-III is managed by the Astrophysical Research Consortium for the Participating Institutions of the SDSS-III Collaboration including the University of Arizona, the Brazilian Participation Group, Brookhaven National Laboratory, University of Cambridge, University of Florida, the French Participation Group, the German Participation Group, the Instituto de Astrofisica de Canarias, the Michigan State/Notre Dame/JINA Participation Group, Johns Hopkins University, Lawrence Berkeley National Laboratory, Max Planck Institute for Astrophysics, New Mexico State University, New York University, Ohio State University, Pennsylvania State University, University of Portsmouth, Princeton University, the Spanish Participation Group, University of Tokyo, University of Utah, Vanderbilt University, University of Virginia, University of Washington, and Yale University.

This publication makes use of data products from the Two Micron All Sky Survey, which is a joint project of the University of Massachusetts and the Infrared Processing and Analysis Center/California Institute of Technology, funded by the National Aeronautics and Space Administration and the National Science Foundation.

{\it Facilities:} \facility{Sloan}, \facility{VLA}, \facility{Hale (TripleSpec)}, 

\bibliography{ems_arXiv.bbl}
\clearpage
\LongTables
%\begin{landscape}
%% LaTeX deluxetable generator for the AASTeX package.
%% Written by Greg Schwarz (5/1/2001).

%% Table generated: Thu May 23 12:59:48 2013

%% Remove the two lines and the last line if you want
%% want to incorporate this table into another LaTex document.
%\documentclass{aastex}
%\begin{document}

%% The values (usually only l,r and c) in the last part of
%% \begin{deluxetable}{} command tell LaTeX how many columns
%% there are and how to align them.
\begin{deluxetable}{cccccccccccccccl}

%% Rotate to a landscape orientation
%\rotate

%% Over-ride the default font size
%% Use 8pt
\tabletypesize{\scriptsize}

%% Use \tablewidth{?pt} to over-ride the default table width.
%% If you are unhappy with the default look at the end of the
%% *.log file to see what the default was set at before adjusting
%% this value.
\tablewidth{0pt}

%% This is the title of the table.
\tablecaption{UKFS Red Quasar Candidates \label{tab:candidates1}}

%% This command over-rides LaTeX's natural table count
%% and replaces it with this number.  LaTeX will increment 
%% all other tables after this table based on this number
%\tablenum{1}

%% The \tablehead gives provides the column headers.  It
%% is currently set up so that the column labels are on the
%% top line and the units surrounded by ()s are in the 
%% bottom line.  You may add more header information by writing
%% another line between these lines. For each column that requries
%% extra information be sure to include a \colhead{text} command
%% and remember to end any extra lines with \\ and include the 
%% correct number of &s.
\tablehead{\colhead{R.A.} & \colhead{Dec} & \colhead{$r$\tablenotemark{a}} & \colhead{$i$\tablenotemark{a}} & \colhead{$Y$\tablenotemark{b}} & \colhead{$J$\tablenotemark{b}} & \colhead{$H$\tablenotemark{b}} & \colhead{$K$\tablenotemark{b}} & \colhead{$J-K$} & \colhead{$r-K$} & \colhead{$S_{\rm 20~cm}$}  & \multicolumn{2}{c}{Spectroscopy} & \colhead{Class} & \colhead{Redshift} & \colhead{Comment} \\ 
\colhead{(J2000)} & \colhead{(J2000)} & \colhead{(mag)} & \colhead{(mag)} & \colhead{(mag)} & \colhead{(mag)} & \colhead{(mag)} & \colhead{(mag)} & \colhead{(mag)} & \colhead{(mag)} & \colhead{(mJy)} & \colhead{Near-IR} & \colhead{Optical} & \colhead{} & \colhead{} & \colhead{}\\
\colhead{(1)} & \colhead{(2)} & \colhead{(3)} & \colhead{(4)} & \colhead{(5)} & \colhead{(6)} & \colhead{(7)} & \colhead{(8)} & \colhead{(9)} & \colhead{(10)} & \colhead{(11)} & \colhead{(12)} & \colhead{(13)}  & \colhead{(14)} & \colhead{(15)} & \colhead{(16)}  } 

%% All data must appear between the \startdata and \enddata commands
\startdata
 00:12:42.70  &  $-$00:45:14.8  &   22.22  &   20.55  &   19.41  &   18.89  &   17.77  &   16.93  &    1.96  &    5.29  &  1.19    & 0911      & \nodata    & QSO        & 0.645    & \\ 
 00:16:00.62  &  $-$00:38:59.9  &   21.72  &   21.17  &   19.27  &   18.83  &   17.63  &   16.28  &    2.55  &    5.43  &  1.32    & 0808      & SDSS3      & QSO        & 2.186   & \\ 
 00:20:54.02  &  $+$00:20:26.6  &   22.29  &   20.92  &   19.15  &   18.69  &   17.78  &   16.83  &    1.86  &    5.46  &  2.45    & 0911      & \nodata    & \nodata    & \nodata   &  \\ 
 00:27:42.22  &  $+$00:05:40.6  &   24.00  &   24.36  &   18.62  &   18.11  &   17.10  &   16.21  &    1.90  &    7.80  &  2.98    & \nodata   & SDSS3      & Galaxy     & 0.415    & \\ 
 00:30:04.96  &  $+$00:25:01.3  &   21.61  &   20.92  &   19.12  &   18.38  &   17.23  &   16.11  &    2.27  &    5.50  &  7.5     & 0808      & \nodata    & QSO        & 2.299    & \\ 
 00:36:59.85  &  $-$01:13:32.3  &   20.23  &   19.72  &   17.65  &   16.58  &   15.11  &   13.56  &    3.03  &    6.67  &  1.92    & \nodata   & Glik07     & QSO        & 0.294    & \\ 
 00:43:45.18  &  $-$00:02:08.0  &   21.91  &   20.58  & $>$20.2  &   19.21  &   17.84  &   16.74  &    2.47  &    5.17  &  2.53    & 0808      & AUS        & ?          & \nodata   & X-ray Source \\ 
 01:20:33.54  &  $+$00:54:59.2  & $>$22.3  & $>$21.3  &   19.14  &   19.19  &   17.97  &   17.00  &    2.18  & $>$5.30  &  3.82    & 0911      & \nodata    & \nodata    & \nodata   &  \\ 
 01:26:15.13  &  $+$00:37:47.5  &   22.56  &   20.97  &   18.91  &   18.88  &   17.40  &   16.77  &    2.11  &    5.79  &  5.98    & 0808      & \nodata    & \nodata    & \nodata   &  \\ 
 01:34:12.71  &  $+$00:03:45.6  &   22.44  &   21.49  &   19.33  &   18.62  &   18.03  &   16.66  &    1.96  &    5.78  &  862.76  & 0911      & \nodata    & QSO        & 0.884    & \\ 
 01:35:18.88  &  $+$00:12:22.4  &   23.25  &   21.67  &   19.40  &   18.95  &   18.04  &   16.94  &    2.02  &    6.32  &  3.5     & 0911      & \nodata    & \nodata    & \nodata   &  \\ 
 01:35:39.42  &  $-$00:43:21.7  &   21.26  &   20.47  &   19.18  &   18.57  &   17.35  &   16.07  &    2.50  &    5.19  &  1.05    & 0808      & SDSS3      & QSO        & 0.442  & \\ 
 01:37:23.46  &  $-$00:05:37.5  &   22.12  &   20.80  &   19.55  &   18.68  &   17.94  &   16.89  &    1.80  &    5.24  &  4.27    & 0911      & \nodata    & \nodata    & \nodata   &  \\ 
 01:37:24.82  &  $+$00:09:50.8  &   22.27  &   22.17  &   19.10  &   18.62  &   17.43  &   16.67  &    1.95  &    5.60  &  8.73    & \nodata   & SDSS3      & Galaxy     & 0.584  & \\ 
 01:41:31.03  &  $+$01:01:01.3  &   21.80  &   21.30  &   19.75  &   19.35  &   18.03  &   16.62  &    2.73  &    5.18  &  3.19    & \nodata   & SDSS3      & QSO        & 2.567   & \\ 
 01:42:16.05  &  $+$01:15:11.7  &   21.76  &   20.74  &   19.74  &   19.28  &   17.76  &   16.72  &    2.56  &    5.04  &  2.04    & 0808      & LRIS       & NLAGN      & 0.685    & \\ 
 01:52:43.18  &  $+$00:20:40.3  &   22.09  &   22.91  &   18.11  &   17.96  &   17.02  &   16.25  &    1.71  &    5.84  &  91.69   & \nodata   & SDSS3      & QSO        & 0.578    & \\ 
 01:54:16.31  &  $-$00:03:05.9  &   22.44  &   20.78  &   19.23  & $>$19.6  &   17.42  &   16.70  & $>$2.90  &    5.74  &  3.0     & \nodata   & \nodata    & \nodata    & \nodata   &  \\ 
 01:54:53.75  &  $+$00:10:18.6  &   21.36  &   20.08  &   18.74  &   18.08  &   17.23  &   16.14  &    1.94  &    5.22  &  1.44    & \nodata   & SDSS3      & Galaxy     & 0.656 & \\ 
 01:56:47.61  &  $-$00:58:07.4  &   20.11  &   19.54  &   18.59  &   18.23  &   16.63  &   14.80  &    3.43  &    5.31  &  6.51    & \nodata   & Glik12     & QSO        & 0.507    & \\ 
 01:58:21.10  &  $+$00:27:52.5  &   22.16  &   20.90  &   19.24  &   18.87  &   17.76  &   16.86  &    2.02  &    5.30  &  1.63    & 1109      & AUS        & QSO        & 1.35     & \\ 
 02:00:23.33  &  $+$00:03:41.8  &   22.17  &   20.87  &   19.36  &   19.07  &   17.93  &   16.90  &    2.17  &    5.27  &  3.86    & \nodata   & AUS        & ?          & 0.866 & \\ 
 02:03:15.86  &  $-$00:14:32.7  &   21.79  &   21.79  &   19.19  &   18.54  &   17.59  &   16.75  &    1.79  &    5.04  &  1.83    & \nodata   & AUS        & Galaxy     & 0.381  & \\ 
 02:03:43.31  &  $-$01:02:24.5  &   21.78  &   20.56  &   19.15  &   18.65  &   17.68  &   16.75  &    1.90  &    5.03  &  1.28    & 0808      & \nodata    & \nodata    & \nodata   &  \\ 
 02:04:45.48  &  $-$01:06:14.5  & $>$22.3  & $>$21.3  &   18.48  &   17.85  &   17.00  &   16.34  &    1.51  & $>$5.96  &  1.09    & 0911      & \nodata    & \nodata    & \nodata   &  \\ 
 02:09:07.50  &  $-$01:05:44.3  &   23.82  &   21.33  &   18.27  &   17.69  &   16.72  &   15.91  &    1.78  &    7.91  &  3.53    & 0808      & SDSS3      & NLAGN      & 0.426  & \\ 
 02:10:38.92  &  $-$00:05:47.5  & $>$22.3  & $>$21.3  &   19.21  &   18.92  &   17.87  &   16.86  &    2.06  & $>$5.44  &  1.78    & \nodata   & SDSS       & Galaxy     & 0.435    & \\ 
 02:11:45.81  &  $+$00:52:13.6  &   22.31  &   21.00  &   19.04  &   18.67  &   17.69  &   16.67  &    2.01  &    5.65  &  3.99    & 0808      & \nodata    & \nodata    & \nodata   &  \\ 
 02:13:45.40  &  $+$00:24:36.6  & $>$22.3  & $>$21.3  &   20.12  &   18.77  &   17.69  &   16.49  &    2.28  & $>$5.81  &  1.16    & 0911      & \nodata    & \nodata    & \nodata   &  \\ 
 02:15:50.90  &  $-$00:42:45.2  &   23.52  &   22.28  &   18.22  &   17.59  &   16.81  &   16.01  &    1.58  &    7.51  &  2.57    & 0808      & SDSS       & Galaxy     & 0.437    & \\ 
 02:16:52.99  &  $-$01:02:33.3  &   22.24  &   21.25  &   19.12  &   18.58  &   17.79  &   16.82  &    1.77  &    5.43  &  1.2     & 0911      & \nodata    & \nodata    & \nodata   &  \\ 
 02:20:03.61  &  $-$00:00:27.3  &   22.27  &   21.17  &   19.29  &   19.09  &   17.80  &   16.90  &    2.19  &    5.38  &  3.72    & \nodata   & AUS        & Galaxy     & 0.884  & \\ 
 02:23:01.17  &  $+$00:10:09.2  &   22.37  &   20.89  &   19.50  &   18.61  &   17.83  &   16.93  &    1.68  &    5.44  &  4.28    & 1109      & \nodata    & QSO        & 0.6?    & \\ 
 02:25:18.77  &  $+$00:34:16.9  &   22.09  &   20.83  &   19.20  &   18.63  &   17.54  &   16.78  &    1.85  &    5.31  &  1.35    & 1109      & \nodata    & ?          & 0.769     & photz \\ 
 02:25:34.32  &  $+$01:08:38.1  &   23.26  &   21.27  &   19.79  &   18.96  &   17.76  &   16.95  &    2.01  &    6.31  &  1.07    & 1109      & \nodata    & ?          & 0.892     & photz \\ 
 02:28:27.41  &  $-$00:02:40.1  &   22.18  &   20.90  &   19.80  &   19.03  &   17.85  &   16.93  &    2.09  &    5.25  &  5.56    & \nodata   & AUS        & ?          & 1.014  & \\ 
 02:33:11.60  &  $+$00:17:21.2  &   22.07  &   20.96  &   19.23  &   18.53  &   17.78  &   16.76  &    1.77  &    5.31  &  1.83    & \nodata   & SDSS3      & Galaxy     & 0.823    & \\ 
 02:47:48.93  &  $-$00:56:28.2  &   22.01  &   20.68  &   19.15  &   18.41  &   17.74  &   16.89  &    1.52  &    5.12  &  3.1     & 0911      & \nodata    & ?          & \nodata   &  \\ 
 02:51:20.04  &  $-$00:23:49.1  &   21.52  &   20.40  &   18.89  &   18.22  &   17.26  &   16.51  &    1.71  &    5.01  &  1.6     & \nodata   & SDSS3      & Galaxy     & 0.635  & \\ 
 03:00:07.15  &  $+$00:53:21.0  &   22.00  &   21.10  &   19.94  &   18.90  &   17.99  &   16.90  &    2.00  &    5.10  &  1.05    & \nodata   & LRIS       & Galaxy     & 0.668    & \\ 
 03:00:37.23  &  $+$01:04:43.6  &   22.05  &   21.01  &   19.00  &   18.66  &   17.61  &   16.92  &    1.73  &    5.12  &  1.64    & \nodata   & \nodata    & \nodata    & \nodata   &  \\ 
 03:01:37.10  &  $+$00:43:21.7  &   21.53  &   20.11  &   18.91  &   18.25  &   17.07  &   16.31  &    1.94  &    5.22  &  2.32    & 0808      & \nodata    & ?          & \nodata   &  \\ 
 03:03:39.85  &  $+$01:06:23.2  &   22.11  &   20.90  &   19.30  &   19.14  &   18.02  &   16.95  &    2.19  &    5.16  &  1.21    & \nodata   & \nodata    & \nodata    & \nodata   &  \\ 
 08:19:06.38  &  $+$04:46:27.2  & $>$22.2  & $>$21.3  &   18.93  &   18.42  &   17.52  &   16.75  &    1.67  & $>$5.45  &  3.35    & \nodata   & \nodata    & \nodata    & \nodata   &  \\ 
 08:20:56.35  &  $+$04:53:40.0  & $>$22.2  & $>$21.3  &   19.55  &   18.90  &   17.89  &   16.98  &    1.92  & $>$5.22  &  1.08    & \nodata   & \nodata    & \nodata    & \nodata   &  \\ 
 08:22:09.47  &  $+$05:23:19.0  & $>$22.2  & $>$21.3  &   19.51  &   18.91  &   17.89  &   16.95  &    1.96  & $>$5.25  &  3.93    & \nodata   & \nodata    & \nodata    & \nodata   &  \\ 
 08:29:30.69  &  $+$04:49:34.3  & $>$22.2  & $>$21.3  &   19.28  &   18.54  &   17.52  &   16.63  &    1.91  & $>$5.57  &  2.06    & 1109      & \nodata    & ?          & \nodata   &  \\ 
 08:30:26.71  &  $+$05:05:21.6  & $>$22.2  & $>$21.3  &   19.11  &   18.49  &   17.57  &   16.53  &    1.97  & $>$5.67  &  1.11    & \nodata   & \nodata    & \nodata    & \nodata   &  \\ 
 08:37:19.72  &  $-$01:15:18.7  & $>$22.2  & $>$21.3  &   19.51  &   18.55  &   17.70  &   16.83  &    1.73  & $>$5.37  &  1.71    & 1109      & \nodata    & ?          & \nodata   &  \\ 
 08:37:59.43  &  $+$05:06:11.1  &   22.08  &   20.87  &   19.11  &   18.48  &   17.57  &   16.78  &    1.70  &    5.31  &  1.60    & \nodata   & \nodata    & \nodata    & \nodata   &  \\ 
 09:06:35.91  &  $+$06:07:32.8  &   22.36  &   21.45  &   19.13  &   18.38  &   17.65  &   16.82  &    1.56  &    5.54  &  1.55    & \nodata   & \nodata    & \nodata    & \nodata & \\ 
 09:08:46.11  &  $+$05:28:43.3  &   21.09  &   20.08  &   19.18  &   18.30  &   17.35  &   16.03  &    2.27  &    5.06  &  5.71    & 0413      & SDSS       & QSO        & 0.801    & \\ 
 09:09:17.33  &  $+$05:41:11.0  &   21.89  &   21.05  &   19.30  & $>$19.6  &   17.69  &   16.65  & $>$2.95  &    5.24  &  1.92    & \nodata   & \nodata    & \nodata    & \nodata   &  \\ 
 09:25:43.94  &  $-$02:57:47.1  & $>$22.2  & $>$21.3  &   19.14  &   18.47  &   17.42  &   16.50  &    1.97  & $>$5.70  &  10.95   & \nodata   & \nodata    & \nodata    & \nodata   &  \\ 
 09:38:52.20  &  $-$02:00:59.5  & $>$22.2  & $>$21.3  &   19.52  &   18.90  &   17.84  &   16.85  &    2.05  & $>$5.35  &  4.11    & \nodata   & \nodata    & \nodata    & \nodata   &  \\ 
 09:42:52.91  &  $-$01:47:21.1  &   22.11  &   21.22  &   19.67  &   19.46  &   18.08  &   16.84  &    2.62  &    5.27  &  1.09    & \nodata   & \nodata    & \nodata    & \nodata   &  \\ 
 09:47:32.60  &  $-$02:15:10.8  &   22.46  &   21.30  &   19.52  &   18.95  &   17.92  &   16.97  &    1.98  &    5.49  &  3.91    & \nodata   & \nodata    & \nodata    & \nodata   &  \\ 
 09:48:14.15  &  $+$06:17:42.2  &   22.83  &   21.27  &   19.54  &   18.87  &   17.87  &   16.91  &    1.96  &    5.92  &  15.0    & \nodata   & \nodata    & \nodata    & \nodata   &  \\ 
 09:48:20.11  &  $-$03:08:47.6  & $>$22.2  & $>$21.3  &   20.09  &   19.17  &   18.05  &   16.98  &    2.19  & $>$5.22  &  4.37    & \nodata   & \nodata    & \nodata    & \nodata   &  \\ 
 09:51:43.30  &  $-$03:09:06.5  & $>$22.2  & $>$21.3  &   19.41  &   18.77  &   18.08  &   16.81  &    1.96  & $>$5.39  &  7.59    & \nodata   & \nodata    & \nodata    & \nodata   &  \\ 
 09:55:12.42  &  $-$03:00:31.7  & $>$22.2  & $>$21.3  &   19.27  &   18.87  &   17.73  &   16.99  &    1.87  & $>$5.21  &  2.95    & \nodata   & \nodata    & \nodata    & \nodata   &  \\ 
 10:01:33.06  &  $-$02:25:16.4  &   22.52  &   20.92  &   19.37  & $>$19.6  &   17.84  &   16.89  & $>$2.71  &    5.63  &  8.29    & \nodata   & \nodata    & \nodata    & \nodata   &  \\ 
 10:17:41.40  &  $+$07:48:00.6  &   22.11  &   20.87  &   18.98  &   18.30  &   17.55  &   16.77  &    1.53  &    5.35  &  2.44    & \nodata   & \nodata    & \nodata    & \nodata   &  \\ 
 12:58:05.74  &  $-$00:14:19.0  & $>$22.2  & $>$21.3  &   19.08  &   18.57  &   17.62  &   16.85  &    1.73  & $>$5.35  &  2.57    & \nodata   & SDSS3      & Galaxy     & 0.602    & \\ 
 13:14:30.15  &  $+$00:13:40.4  &   22.06  &   20.83  &   19.34  & $>$19.6  & $>$18.8  &   16.74  & $>$2.86  &    5.31  &  2.25    & \nodata   & \nodata    & \nodata    & \nodata   &  \\ 
 13:19:10.70  &  $+$00:09:55.6  &   21.14  &   20.01  &   20.20  &   18.71  &   17.16  &   16.04  &    2.67  &    5.10  &  47.45   & 0309      & \nodata    & ?          & \nodata   &  \\ 
 13:50:33.48  &  $-$00:24:42.3  &   21.80  &   21.08  &   18.61  &   18.19  &   17.19  &   16.34  &    1.85  &    5.46  &  1.19    & \nodata   & SDSS       & Galaxy     & 0.36     & \\ 
 14:06:43.69  &  $-$00:11:47.1  &   23.16  &   24.02  &   18.93  &   18.48  &   17.53  &   16.83  &    1.65  &    6.32  &  1.58    & \nodata   & \nodata    & \nodata    & \nodata   &  \\ 
 15:33:47.27  &  $+$05:41:50.9  &   22.40  &   21.69  &   19.07  &   18.59  &   18.08  &   16.99  &    1.60  &    5.42  &  1.54    & \nodata   & \nodata    & \nodata    & \nodata   &  \\ 
 15:42:46.51  &  $+$05:36:04.7  &   22.12  &   20.97  &   19.49  &   18.88  &   17.88  &   16.91  &    1.98  &    5.22  &  2.59    & 0309      & \nodata    & ?          & \nodata   &  \\ 
 15:44:01.86  &  $+$05:34:51.8  &   21.85  &   20.84  &   19.33  &   18.82  &   17.79  &   16.81  &    2.01  &    5.03  &  1.14    & \nodata   & \nodata    & \nodata    & \nodata   &  \\ 
 22:20:55.61  &  $-$00:50:13.7  &   22.18  &   24.34  &   19.27  &   18.53  &   17.56  &   16.73  &    1.81  &    5.45  &  19.9    & 1109      & \nodata    & ?          & \nodata   &  \\ 
 22:22:09.79  &  $-$00:44:21.3  &   22.59  &   21.05  &   19.52  &   18.76  &   17.75  &   16.72  &    2.04  &    5.87  &  3.81    & 0808      & \nodata    & ?          & \nodata   &  \\ 
 22:25:17.81  &  $-$00:17:18.0  &   22.32  &   20.91  &   19.24  &   18.56  &   17.97  &   16.81  &    1.75  &    5.51  &  1.52    & 1109      & \nodata    & ?          & \nodata   &  \\ 
 22:30:59.65  &  $+$00:22:09.1  &   22.75  &   21.38  &   19.75  &   19.01  &   17.99  &   17.00  &    2.02  &    5.75  &  6.34    & 0911      & \nodata    & QSO        & 0.65     & \\ 
 22:38:19.00  &  $+$01:05:09.3  & $>$22.2  & $>$21.3  & $>$20.5  &   18.98  &   17.87  &   17.00  &    1.98  & $>$5.30  &  65.64   & \nodata   & \nodata    & \nodata    & \nodata   &  \\ 
 22:42:39.86  &  $-$00:38:47.6  &   21.76  &   20.90  &   19.52  &   18.92  &   17.73  &   16.25  &    2.67  &    5.51  &  1.28    & 0808      & AUS        & NLAGN      & 0.615    & \\ 
 22:43:38.05  &  $+$00:17:50.0  & $>$22.2  & $>$21.3  & $>$20.2  & $>$19.6  &   18.21  &   16.78  & $>$2.82  & $>$5.52  &  80.09   & 0911      & \nodata    & ?          & \nodata   &  \\ 
 22:45:48.33  &  $-$00:14:56.1  &   22.35  &   20.63  &   19.49  &   18.82  &   17.81  &   16.96  &    1.86  &    5.39  &  5.45    & \nodata   & LRIS       &  Galaxy    &  0.625    & \\ 
 22:50:09.90  &  $+$00:10:41.0  &   23.57  &   22.15  &   18.99  &   18.43  &   17.74  &   16.68  &    1.74  &    6.89  &  4.6     & 0808      & 2SLAQ      & Galaxy     & 0.645   & \\ 
 22:54:23.44  &  $+$00:27:58.4  &   22.21  &   20.84  &   19.33  &   18.82  &   17.75  &   16.94  &    1.88  &    5.27  &  3.35    & 1109      & AUS        & ?          & 0.807  & \\ 
 23:34:52.43  &  $-$00:24:22.6  &   20.94  &   20.26  &   18.82  &   18.45  &   17.14  &   15.93  &    2.52  &    5.01  &  1.15    & 1109      & AUS        & QSO        & 0.557  & \\ 
 23:34:54.44  &  $+$00:16:10.5  &   22.20  &   20.40  &   19.49  &   18.64  &   17.75  &   16.84  &    1.80  &    5.36  &  4.22    & 0911      & \nodata    & ?          & \nodata   &  \\ 
 23:36:33.52  &  $-$00:49:57.3  &   21.81  &   21.00  &   19.39  &   18.69  &   17.54  &   16.78  &    1.91  &    5.03  &  1.24    & 0911      & \nodata    & ?          & \nodata   & merger \\ 
 23:37:23.59  &  $-$00:58:11.3  &   21.83  &   21.08  &   19.45  &   18.71  &   17.76  &   16.77  &    1.94  &    5.07  &  40.57   & 0808      & \nodata    & ?          & \nodata   &  \\ 
 23:37:56.73  &  $-$00:32:22.6  &   22.10  &   20.64  &   19.15  &   18.59  &   17.64  &   16.57  &    2.03  &    5.54  &  4.41    & 0808      & AUS        & ?          & 0.719  & \\ 
 23:40:15.28  &  $-$00:10:33.7  &   21.31  &   20.53  &   19.00  &   18.09  &   17.05  &   16.08  &    2.01  &    5.23  &  5.33    & 1109      & AUS        & ?          & \nodata   &  \\ 
\enddata

%% Include any \tablenotetext{key}{text}, \tablerefs{ref list},
%% or \tablecomments{text} between the \enddata and 
%% \end{deluxetable} commands

%% General table comment marker
\tablecomments{Optical spectra:
SDSS = SDSS DR7 spectrum;
SDSS3 = SDSS DR9 spectrum;
AUS = redshift from AUS (Croom et al., in prep);
LRIS = Keck LRIS 2012 October; 
2SLAQ = 2dF-SDSS LRG and QSO survey \citep{Cannon06}.
Infrared Spectra:
0808 = TripleSpec 2008 August;
0309 = TripleSpec 2009 March;
1109 = TripleSpec 2009 November;
0911 = TripleSpec 2011 September;
0413 = TripleSpec 2013 April.}

%% General table references marker
\tablerefs{Glik07 = F2M Quasar from \citet{Glikman07};
Glik12 = F2M Quasar from \citet{Glikman12}; 
H05 = Photometric redshifts from \citet{Hsieh05};
H09 = \citet{Hwang09}.}

\end{deluxetable}
%\end{document}

\clearpage
%\end{landscape}
\clearpage
%% LaTeX deluxetable generator for the AASTeX package.
%% Written by Greg Schwarz (5/1/2001).

%% Table generated: Thu Aug 30 11:57:58 2012

%% Remove the two lines and the last line if you want
%% want to incorporate this table into another LaTex document.
%\documentclass{aastex}
%\begin{document}

%% The values (usually only l,r and c) in the last part of
%% \begin{deluxetable}{} command tell LaTeX how many columns
%% there are and how to align them.
\begin{deluxetable}{cccccccccc}

%% Keep a portrait orientation
%\rotate

%% Over-ride the default font size
%% Use Default (12pt)
\tabletypesize{\small}

%% Use \tablewidth{?pt} to over-ride the default table width.
%% If you are unhappy with the default look at the end of the
%% *.log file to see what the default was set at before adjusting
%% this value.
 \tablewidth{0pt}
 
%% This is the title of the table.
\tablecaption{Reddening Properties of UKFS red quasars \label{tab:ebv}}

%% This command over-rides LaTeX's natural table count
%% and replaces it with this number.  LaTeX will increment 
%% all other tables after this table based on this number
%\tablenum{3}

%% The \tablehead gives provides the column headers.  It
%% is currently set up so that the column labels are on the
%% top line and the units surrounded by ()s are in the 
%% bottom line.  You may add more header information by writing
%% another line between these lines. For each column that requries
%% extra information be sure to include a \colhead{text} command
%% and remember to end any extra lines with \\ and include the 
%% correct number of &s.
\tablehead{
\colhead{} & \colhead{} & \colhead{} & \colhead{} & \colhead{} & \multicolumn{4}{c}{E(B-V)} & \colhead{de-reddened}\\
\colhead{Name} & \colhead{R.A.} & \colhead{Dec} & \colhead{Redshift} & \colhead{$K_s$} & \colhead{Photom} & \colhead{Optical} & \colhead{Near-IR} & \colhead{Combined} & \colhead{$M_K$} \\ 
\colhead{} & \colhead{(J2000)} & \colhead{(J2000)} & \colhead{} & \colhead{(mag)} & \colhead{(mag)} & \colhead{(mag)} & \colhead{(mag)} & \colhead{(mag)} & \colhead{(mag)} } 

%% All data must appear between the \startdata and \enddata commands
\startdata
UKFS 0141+0101   & 01:41:31.03 & $+$01:01:01.3 & 2.567 & 16.62 &  0.20  &   0.15 &   ...  & ... & $-30.53$ \\ 
UKFS 0030+0025   & 00:30:04.96 & $+$00:25:01.3 & 2.299 & 16.11 &  0.32  &   ...  &   0.21 & ... & $-30.98$\\ 
UKFS 0016$-$0038 & 00:16:00.62 & $-$00:38:59.9 & 2.186 & 16.28 &  0.30  &   0.18 &   0.02 & 0.30 & $-30.59$\\ 
UKFS 0158+0027   & 01:58:21.10 & $+$00:27:52.5 & 1.350 & 16.86 &  0.33  &   ...  &   0.10 & ... & $-28.50$\\ 
 &                         &                            &           &            & 3.90\tablenotemark{a} &        &               &      & $-31.94$\tablenotemark{a}\\
UKFS 0134+0003   & 01:34:12.71 & $+$00:03:45.6 & 0.884 & 16.66 &  0.53  &   ...  &   0.14 & ... & $-27.60$\\ 
 &                         &                            &           &            & 1.17\tablenotemark{a} &        &               &      & $-26.84$\tablenotemark{a}\\
UKFS 0908+0528   & 09:08:46.11 & $+$05:28:43.3 & 0.801 & 16.03 &  0.63  &   0.67 &   0.17 & 0.58 & $-28.01$\\ 
 &                         &                            &           &            & 2.91\tablenotemark{a} &        &               &      & $-29.50$\tablenotemark{a}\\
UKFS 2230+0022   & 22:30:59.65 & $+$00:22:09.1 & 0.65  & 17.00 &  0.59  &   ...  &$-$0.34 & ... & $-26.39$\\
 &                         &                            &           &            & 1.01\tablenotemark{a} &        &               &      & $-25.28$\tablenotemark{a}\\
UKFS 0012$-$0045 & 00:12:42.70 & $-$00:45:14.8 & 0.645 & 16.93 &  0.60  &   ...  &   0.13 & ... & $-26.44$\\ 
 &                         &                            &           &            & 2.02\tablenotemark{a} &        &               &      & $-25.63$\tablenotemark{a}\\
UKFS 0223+0010   & 02:23:01.17 & $+$00:10:09.2 & 0.6   & 16.93 &  0.67  &   ...  &   0.18 & ... & $-26.29$\\ 
 &                         &                            &           &            & 1.06\tablenotemark{a} &        &               &      & $-24.89$\tablenotemark{a}\\
UKFS 0152+0020   & 01:52:43.18 & $+$00:20:40.3 & 0.578 & 16.25 &  0.19  &   0.42 &    ... & ... & $-26.52$\\ 
UKFS 2334$-$0024 & 23:34:52.43 & $-$00:24:22.6 & 0.557 & 15.93 &  0.78  &   ...  &   0.34 & ... & $-27.14$\\ 
UKFS 0156$-$0058 & 01:56:47.61 & $-$00:58:07.4 & 0.507 & 14.80 &  0.75  &   0.50 &   1.50 & 0.89 & $-27.98$\\ 
UKFS 0135$-$0043 & 01:35:39.42 & $-$00:43:21.7 & 0.442 & 16.07 &  0.61  &   0.69 &$-$0.22 & 0.70 &$-26.24$\\ 
UKFS 0036$-$0113 & 00:36:59.85 & $-$01:13:32.3 & 0.294 & 13.56 &  1.09  &   1.07 &   ...  & ... & $-27.90$\\ 
\enddata

%% Include any \tablenotetext{key}{text}, \tablerefs{ref list},
%% or \tablecomments{text} between the \enddata and 
%% \end{deluxetable} commands
\tablenotetext{a}{These values are based on the two-component quasar plus host galaxy fits described in Section \ref{sec:ebv}.}
%% No \tablecomments indicated

%% No \tablerefs indicated

\end{deluxetable}
%\end{document}

\clearpage

\end{document}